%% file: schuessler_susy07.tex
\documentclass[epj]{svjour}

\usepackage{graphicx}
\usepackage{fancyhdr}

\usepackage{amsmath,amssymb,stmaryrd,mathrsfs}
\usepackage{psfrag}

\newcommand{\ct}{\mathcal{T}}
\newcommand{\vp}{\varphi}
\newcommand{\si}{\mathrm{i}}
\newcommand{\sI}{\mathrm{I}}
\newcommand{\sd}{\mathrm{d}}
\newcommand{\sR}{\mathrm{R}}
\newcommand{\se}{\mathrm{e}}
\newcommand{\sm}{\mathrm{m}}
\newcommand{\sta}{\tilde{t}_1}
\newcommand{\stb}{\tilde{t}_2}
\newcommand{\sba}{\tilde{b}_1}
\newcommand{\sbb}{\tilde{b}_2}
\newcommand{\asta}{\tilde{t}^*_1}

\newcommand{\asba}{\tilde{b}^*_1}

\def\makeheadbox{{
 }}
\def\mytitle{My title} 
\def\myauthors{My name}  
\def\mytype{My type of session}
\def\mysession{My session}

\def\mytitle{Unitarity constraints on MSSM trilinear couplings}
\def\myauthors{Alexander Schuessler}    
\def\mytype{Unitarity constraints on trilinear couplings in the MSSM}    
\def\mysession{Colliders - SUSY Phenomenology}

\begin{document}
\title{Unitarity constraints on MSSM trilinear couplings}
\author{Alexander Schuessler \inst{1}\inst{2}
\thanks{\emph{Email:} alesch@tkm.uni-karlsruhe.de}
 \and
 Dieter Zeppenfeld\inst{3}
}                     
\institute{Institut f\"ur Nanotechnologie, 
Forschungszentrum Karlsruhe, 76021 Karlsruhe, Germany
\and Institut f\"ur Theorie der Kondensierten Materie, 
Universit\"at Karlsruhe,
76128 Karlsruhe, Germany
\and Institut f\"ur Theoretische Physik, 
Universit\"at Karlsruhe,
76128 Karlsruhe, Germany
}
%
\date{}
\abstract{For MSSM phenomenology, soft SUSY breaking dimension-three 
operators are important, in particular the couplings between Higgs 
bosons and squarks. In scattering processes, perturbative unitarity 
is violated at modest center-of-mass energy if these couplings are much 
larger than the masses of the scalar particles involved. Assuming 
perturbative unitarity, constraints on the trilinear couplings can be 
determined using a computer program that we have developed.
} 
\maketitle
\section{Introduction}
\label{intro}
For phenomenology in the Minimal Supersymmetric Standard Model (MSSM), 
the soft supersymmetry \linebreak (SUSY) breaking dimension-three trilinear 
couplings between Higgs bosons and squarks play an important role. They
determine the mass splitting in the stop and sbottom sector and 
can have a profound influence on Higgs physics. Unfortunately, these
SUSY breaking couplings are essentially free parameters. Thus, 
any constraints that can be placed by theoretical considerations 
are useful. Such constraints can be obtained, for example, by considering
the symmetry breaking of the scalar sector and requiring that there are
no color or charge breaking minima~\cite{Casas}. Requiring that the MSSM
allows for a perturbation theory treatment, we here consider a
different approach: perturbative unitarity in $2\to 2$ scattering
of scalar particles limits the size of the trilinear couplings.

The approach is similar to the consideration of longitudinal weak boson
scattering in the Standard Model (SM), where the absence of perturbative
unitarity violation at high energies provides an upper bound on the
Higgs quartic coupling and thus on the Higgs boson mass~\cite{LQT,HK}. 
For trilinear couplings in the MSSM scalar sector, these unitarity violations
arise at intermediate energies, somewhat above pair production
thresholds, if trilinear couplings are chosen much larger than the masses
of the scalar particles in the scattering process. 
The approach is applied to the trilinear couplings between
Higgs bosons and 3rd generation squarks, given by 
\begin{align*}
 {\cal L}_{\mathrm{tri}} =
 -\lambda_b A_b H_d \tilde{Q}_L \tilde{b}_R^\dagger\!
 -\lambda_t A_t H_u \tilde{Q}_L \tilde{t}_R^\dagger\!
 +\mathrm{h.c.}
\end{align*}
which is part of the MSSM soft SUSY  breaking potential. 
Here $\tilde{Q}_L$ is the $\mathit{SU}\!(2)_L$-doublet of 3rd generation squarks,  $\tilde{b}_R^\dagger$ and $\tilde{t}_R^\dagger$ denote the right-handed
sbottom and stop singlet fields, and $H_d$ and $H_u$ are the Higgs boson
doublet fields. We want to limit the dimensionful
parameters which multiply the Yukawa couplings 
$\lambda_b\!=\!m_b/v\cos\beta$ and $\lambda_t\!=\!m_t/v\sin\beta$, and $\mu$, 
the Higgs mixing parameter.
%
%
%
\section{Perturbative unitarity for scalars}
\label{sec:unitarity}
The starting point is the unitarity of the $S$ matrix,
\begin{align*}
S^\dagger S =1\,, \;\text{ or equivalently }\; -\si\left(T-T^\dagger \right)=T^\dagger T \,,
\end{align*}
for the transition operator $T$ with $S\!=\!1\!+\!\si\,T\!$.
To evaluate this equation one usually restricts to $2\!\rightarrow\! 2$
scattering (which is a good approximation in perturbation theory)
and then uses angular-momentum conservation and symmetries of the model 
to partially diagonalize $T$. In our analysis we additionally restrict
ourselves to scalar fields.  
Let $\langle f|T|i\rangle$ denote the transition matrix elements with
initial state $|i\rangle$ and final state $|f\rangle$, and let
$\hat{\ct}_{fi}$ be the matrix element obtained from Feynman diagrams in
momentum representation, 
\begin{align*}
(2\pi)^4\,\delta^{(4)}(P_i-P_f)\;\hat{\ct}_{fi}(\sqrt{s},\cos\theta)=
\langle f|T|i\rangle\,,
\end{align*}
evaluated in the center-of-mass system, where $\sqrt{s}$ is the total
energy and $\theta$ is the scattering angle.
This transition is defined by the (ordered) particle content of the
two states, as well as $\sqrt{s}$ and $\theta$, and depends on the masses
$m_{lk}$ of the particles in state $l\!=\! i,f$ via the functions
$\lambda_l\! =\! \lambda(s,m_{l1}^2,m_{l2}^2)$, 
$\lambda(x,y,z)\!=\!x^2\!+\!y^2\!+\!z^2 \!-\!2xy\!-\!2yz\!-\!2zx$.
The dependence on the scattering angle $\theta$  is eliminated
 by projection onto partial
waves of total angular momentum $J=\!0,1,2,\ldots$
\begin{align*}
&\ct_{fi}^J =
\frac{1}{2}\,\frac{\lambda^{1/4}_f\lambda^{1/4}_i}{16\pi\,s}
\int_{-1}^1\sd\!\cos\theta\,
\hat{\ct}_{fi}(\sqrt{s},\cos\theta) \,P_J(\cos\theta) \,.
 \end{align*}
The $P_J$ are the Legendre polynomials. The factor 1/2 is a standard
convention and leads to the factor 1/2 in
Eq. \eqref{eq:unitary}. 
In this normalization, extra factors of $1/\sqrt{2}$ have to be included
for each state with two identical particles. Higher partial waves usually
give smaller amplitudes, so only $J\!=\!0,1$ amplitudes have to be
considered in practice. 
The unitarity condition now reads
\begin{align}
\label{eq:unitary}
\frac{1}{2\si}\left(\ct_{fi}^J-\ct_{if}^{J*}\right)\cong
 \sum_h \ct_{hf}^{J*}\,\ct_{hi}^J\,.
\end{align}
The sum is taken over intermediate states. Restriction to only relevant
and two-particle scalar states in the sum slightly underestimates the
right-hand side and leads to conservative bounds.

The 'true', physical  matrix $\ct_{fi}^J$ is normal and can therefore be
diagonalized. The same holds for the Born amplitude, which we use instead.
The diagonalized matrix $\tilde{\ct}_{fi}^J$ and thus the eigenvalues
satisfy 
\begin{align}
\label{eq:ev}
\sI \sm\, \tilde{\ct}_{ii}^J\cong |\tilde{\ct}_{ii}^{J}|^2 \,.
\end{align}
The 'true' eigenvalues (for any given energy $\sqrt{s}$) must lie on the  
circle (called Argand diagram) given by \eqref{eq:ev}: 
$y\!=\!x^2\!+y^2$ for $x=\!\sR\se\, \tilde{\ct}^J_{ii}$ and 
$y=\!\sI\sm\, \tilde{\ct}^J_{ii}$ which implies
$|x|\!\leq\!1/2$. For the Born approximation, the phases of the fields can be
chosen such that all $2\!\rightarrow\! 2$ 
amplitudes are (nearly) real, if $\mathit{CP}$ (nearly) holds.
Approximating $x$ by the corresponding Born amplitude yields the desired
unitarity bound. 
The circle restricts $|x|\!\leq\! 1/2$, which is the unitarity 
bound generally used in the literature.
A Born value of $x\!=1/2, y\!=\!0$ 
needs at least\footnote{'At least' 
means that this would be the minimal case where the correction 
directly hits the nearest circle point.} 
a correction of $\sqrt{2}\!-\!1\!\approx\! 41\%$ to
become unitary.  Such large corrections indicate a breakdown of
perturbation theory. In addition to the perturbative unitarity bound of
$|x|\!\leq\!1/2$ we therefore also consider $|x|\!\leq\!1/6$ 
as a condition for which the Born amplitude remains sufficiently small to
trust perturbation theory. 
\subsection{A toy model}
\label{sec:amplitudes}
{%
\begin{figure}
\psfrag{f1}[rc][rb]{\small{$\vp_1$}}
\psfrag{f2}[rc][rb]{\small{$\vp_2$}}
\psfrag{f3}[lc][lb]{\small{$\vp_3$}}
\psfrag{f4}[lc][lb]{\small{$\vp_4$}}
\psfrag{f5}[rc][rb]{\small{$\vp_5$}}
\psfrag{f6}[b][b]{\small{$\vp_5$}}
\psfrag{s}[c][c]{\small{$(s)$}}
\psfrag{t}[c][c]{\small{$(t)$}}
\psfrag{u}[c][c]{\small{$(u)$}}
\centerline{\includegraphics[width=0.45\textwidth,angle=0]{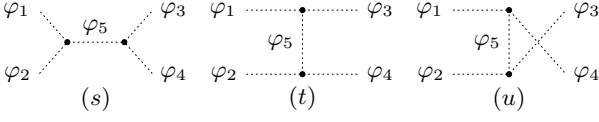}}
\caption{Scalar $2\!\rightarrow\! 2$ tree level scattering diagrams.}
\label{fig:Feynman}  
\end{figure}
}%
Figure \ref{fig:Feynman} shows generic tree level Feynman graphs for
$\vp_1 \vp_2 \rightarrow \vp_3\vp_4$ 
scattering with trilinear couplings of (possibly different) scalar fields 
$\vp_l$ ($l\!=\!1,\!..,5$).
The corresponding amplitudes are $A^2/(q^2\!-m_5^2)$, where $q^2\!=\!s,t,u$ is
one of the Mandelstam variables, and $A$ stands for the trilinear
couplings. 
$t$ and $u$ depend on $\sqrt{s}$, $\cos\theta$, and the masses $m_l$ of
the exterior particles. Projecting onto partial waves, the 
amplitude for $J\!=\!0$ has a structure roughly like: 
\begin{align}\label{eq:toy}
\ct_{fi}^{J=0} \sim \frac{1}{16\pi}\,
\frac{\lambda^{1/4}_f\lambda^{1/4}_i}{s}\,\frac{A^2}{\max\{s,m^2_5\}}\,,
\end{align}
where a factor of 2 has been assumed to account for the partial wave
projection. The second factor is smaller than 1 
 and the third factor becomes large if both the scattering energy
 $\sqrt{s}$ and the mass of the internal particle are small compared to
 the couplings $A$. Highest values are usually found for energies near
 the kinematic threshold, i.e. at energies where the model should work
 properly, in contrast to weak boson scattering in the SM~\cite{LQT,HK}. 
\subsection{Handling poles}
\label{sec:handlingpoles}
Clearly the Born amplitudes are not sufficient to describe scattering
processes where intermediate particles become on-shell. 
In the $s$ channel in Figure \ref{fig:Feynman} this happens when 
$\sqrt{s}\!=\!m_5$. 
Born amplitudes only will be used for unitarity considerations and
$s$-channel poles are cut out by the condition 
\begin{align}
\label{eq:schannel}
|\sqrt{s}-m|^2 > a\,m\,\Gamma(Q\!=\!b\,m)\,. 
\end{align}
Here $a, b\!\gtrsim \!1$ are (suitably chosen) constants and the 
'running width' $\Gamma(Q)$ of the internal particle $\vp$ at
energy $Q$ is approximated by the decay width via replacing its mass $m$
with the energy $Q$ in the phase space factor. This condition
\eqref{eq:schannel} has to be fulfilled for all internal particles
appearing in the $s$ channel. If this is not the case, the amplitude is
set to zero, as well as the irreducible part of $\ct_{fi}^{J=0}$ this 
process is in, because of possible destructive interference of matrix elements.

A width cannot be included in the Born propagator for two reasons:
First, $\ct_{fi}$ is no longer diagonalizable (at this level of
approximation). Second, our $\Gamma(\sqrt{s})$ grows (linearly
for large $\sqrt{s}$) with $\sqrt{s}$, which is not a good approximation
to the propagator as $\sqrt{s}\!\gg \!m$. 

The internal particle in the $u$ channel of Figure \ref{fig:Feynman} can
also become on-shell for certain combinations of masses. This occurs
e.g. if $\vp_1$ can decay into $\vp_4, \vp_5$ and $\vp_2,\vp_5$ can fuse
to $\vp_3$ (MSSM example: $\stb\sta\rightarrow\stb\sta$ with $u$ channel
$h^0$ exchange when $m_{\stb}\!>\!m_{\sta}\!+m_h$). 
One obtains another possibility by switching
labels $1\!\leftrightarrow\!2$ and $3\!\leftrightarrow\!4$ or two similar
conditions for the $t$ channel by exchange of
$3\!\leftrightarrow\!4$. The first case has the condition: 
\begin{align}
\label{eq:uchannelexample}
c\,m_1 \geq m_4+m_5 \; \wedge\; m_2+m_5 \leq c\, m_3\,. 
\end{align}
with some suitably chosen constant $c\gtrsim 1$. 
Amplitudes where a condition
like in \eqref{eq:uchannelexample} is fulfilled cannot be computed
because the internal particle becomes on-shell for some value of the
scattering angle. The constants $a,b,c$ are chosen
larger than one because in proximity of a pole one encounters unphysical
enhancements of the amplitude in the pure Born approximation. 
Still, some enhancement can appear in special cases.  

If some Born matrix elements cannot be calculated because of a $t$ or
$u$ channel pole, the tree level matrix $\ct_{fi}^J$ cannot be
diagonalized. The solution is a partial diagonalization. Assuming time
reflection invariance, we write the left-hand side of \eqref{eq:unitary} as
$\sI \sm \,\ct_{fi}^J$. Define the set $B$ 
of all kinematically accessible states at given $\sqrt{s}$ from an
irreducible part of $\ct_{fi}^J$ and the set $C\!\subset\!B$ such that
$h,l\!\in\!C$ satisfy a $t$ or $u$ channel pole condition 
for $h\!\rightarrow\!l$.
Then, for states $f, i\! \in\! B\backslash C$ equation 
\eqref{eq:unitary} becomes   
\begin{align*}
\sI \sm \,\ct_{fi}^J\,\cong\!\sum_{h\in B\backslash C} \ct_{hf}^{J*}\,\ct_{hi}^J\,+
\sum_{h\in C} \ct_{hf}^{J*}\,\ct_{hi}^J\,.
\end{align*}
We diagonalize the sub-matrix $(\ct_{fi}^J)|_{f,i\in B\backslash C}$ of all \linebreak states
accessible at a given energy $\sqrt{s}$ with a unitary matrix $U$ and
write the diagonalized part as $\tilde{\ct}$ to obtain: 
\begin{align*}
\sI \sm \,\tilde{\ct}_{ii}^J\cong |\tilde{\ct}_{ii}^J|^2 + \sum_{h\in C}
 \left|(\ct^J \big|_{C \times B\backslash C}\,U^{-1})_{ih}\right|^2\!.
\end{align*}
The second term on the right-hand side (denoted by $R^2$) is
positive. At low energies where only a few states are kinematically
allowed, $R^2$ typically vanishes.
The circle equation $y\!=\!x^2\!+y^2$ now holds for $y\!=\!\sI\sm\,
\tilde{\ct}_{ii}^J$ and  
\begin{align}
\label{eq:defx}
|x|= (\,(\sR\se\, \tilde{\ct}_{ii}^J)^2 +R^2)^{1/2}\,.
\end{align}
\section{MSSM constraints}
\label{sec:MSSM}
The MSSM has a lot of ($>\!100$) parameters, many of which are constrained 
by measurement of the properties of SM particles 
and (nearly) conserved symmetries.
Here we focus on some of the SUSY breaking parameters, namely the interplay
of mass terms for the scalar fields and the 
dimensionful $A$ parameters of the trilinear couplings, and also $\mu$. 
These parameters 
are constrained by lower mass bounds for the sparticles and
the Higgs bosons~\cite{neutralHiggs}.
Their phases are bounded by the absence of large $\mathit{CP}$ violation.
In addition, the fields must be in the observed minima of the 
potential and have positive masses after electroweak symmetry breaking. 
Complex, strong constraints from not being in charge and color 
breaking minima were derived in~\cite{Casas}.  
The new constraints from perturbative 
unitarity should be considered in addition.
\subsection{Particles, parameters and calculation}
\label{sec:relpar}
In the MSSM large trilinear couplings may appear between 3rd
generation squarks and the Higgs bosons and the longitudinal degrees of
freedom of $W^\pm$ and $Z^0$.  
To work in the scalar sector only, the Goldstone equivalence theorem and 
the Feynman $R_{\xi = 1}$-gauge are used, such that the Goldstone 
bosons $G^\pm$, $G^0$  represent the longitudinal polarizations 
of $W^\pm$, $Z^0$. The trilinear coupling strengths are given by vertex 
factors like
\begin{align*}
V(G^0\!,\tilde{t}_L,\tilde{t}_R)=g' \, (A_t-\mu^*\cot \beta)\, m_t/2m_W\,.
\end{align*}
Other stop vertices with Higgs bosons are similar and contain 
combinations of $A_t$ and $\mu$  (in decoupling scenarios mainly  
$X_t\!=\!A_t\!-\mu^*\cot \beta$). Likewise two sbottom Higgs 
boson couplings grow with 
$A_b$ and $\mu$, but are suppressed compared to the stop case, 
because they contain the factor $m_b/2m_W$.
Some mixed stop-sbottom couplings also behave like the pure stop case, 
e.g. the vertex $V(G^+\!, \tilde{b}_L, \tilde{t}_R)\!=\! \si\, 
g'\,X_t\, m_t/\sqrt{2}m_W$.

Relevant for our analysis are  
the scalar Higgs bos\-ons $h^0\!$, $H^0\!$, $A^0\!$, $H^\pm\!$, $G^0\!$, and 
$G^\pm$ and the heavy squark mass eigenstates $\sba$, $\sbb$, $\sta$,
and $\stb$. In addition to SM parameters, results will depend on
the squark mass parameters $M^2_{\tilde{Q}}$, $M^2_{\tilde{b}}$, and
$M^2_{\tilde{t}}$, the mass of the pseudo-scalar Higgs boson $m_{\!A}$, the
Higgs mixing parameter $\mu$, the ratio of Higgs bosons VEVs
$\tan\beta$  and the trilinear coupling parameters  
$A_b$ and $A_t$.
Neglecting any $\mathit{CP}$-violating effects, these parameters
are chosen real in the following. 

With the scalar fields listed above, two-particle \linebreak
 states are formed. 
One uses charge, color, and the non-existence of three-squark-vertices 
 to form 15 
independent blocks in the scattering matrix. 
The biggest block (charge 0, color singlet) 
contains 21 states, at high enough energy, and usually supplies the
largest eigenvalue.    
A Monte Carlo search over the scattering energy $\sqrt{s}$ is used to find 
the largest eigenvalue of the scattering matrix, which is obtained by
numerically diagonalization.  
We use, by default, $\overline{\text{MS}}$  quark masses (at 120 GeV) to
calculate the squark vertices and masses. 

Large trilinear couplings can lead to strong loop suppressions of SUSY Higgs
masses~\cite{2loopHM} in regions where our analysis indicates 
a breakdown of perturbation theory. In such cases we ignore clashes with
LEP \linebreak 
bounds~\cite{neutralHiggs} and conservatively use
$m_{h^0}\!=\!120$~GeV instead.   
\subsection{Relevance of the method for the MSSM}
\label{sec:relevance}
Because only heavy squarks and Higgs bosons were used, no parameters for
 charginos and other squarks or sleptons were needed for the calculation. 
The results are also practically independent of these other parameters. 
%
%
%
The maximum mixing case, $X_t\!\gtrsim$ 2 times the sfer\-mion mass scale,
is often used as a benchmark scenario.
An important bound for these cases often used in literature derives 
from \cite{Frere} (see e.g. Eq.(5) in~\cite{Casas}):
\begin{subequations}
\label{eq:Avac}
\begin{align}
\label{eq:Atvac}
A_t^2 &\leq 3 \;\big(\, M_{\tilde{Q}}^2 + M_{\tilde{t}}^2+ m_{H_u}^2\!+|\mu|^2\,\big)\,,\\
\label{eq:Abvac}
A_b^2 &\leq 3 \;\big(\,M_{\tilde{Q}}^2 + M_{\tilde{b}}^2+ m_{H_d}^2\!+|\mu|^2\,\big)\,,
\end{align}
\end{subequations}
with $m_{H_{u\,\!(d)}}^2\!+|\mu|^2= m_A^2 \cos^2\!\beta\, 
(\sin^2\!\beta)\pm1/2\, m_Z^2 \cos 2\beta$.
For large $\tan\beta$ this means $|A_t|^2 \!\lesssim\! 3\,
(M_{\tilde{Q}}^2\! +\! M_{\tilde{t}}^2\!-\!m_Z^2/2)$ and $|A_b|^2\!
\lesssim 3\,(M_{\tilde{Q}}^2\! + M_{\tilde{b}}^2\!+
m_A^2\!+m_Z^2/2)$.
The $A_t$ bound in \eqref{eq:Atvac} is mostly determined by 
$M_{\tilde{Q}}^2 + M_{\tilde{t}}^2$. The unitarity bounds instead depend 
on the masses of the lightest particles with large trilinear couplings, 
which therefore can be stronger in the case of exceptionally light 
scalars while retaining a large value for 
$M_{\tilde{Q}}^2 + M_{\tilde{t}}^2$.
\subsection{Examples for unitarity bounds}
\label{sec:examples}
\begin{figure}
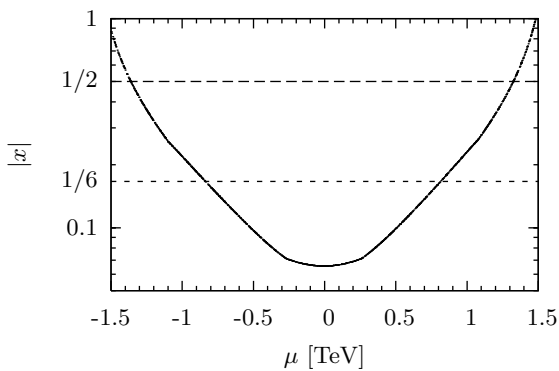

\include{schuessler_susy07fig2}
\vspace{-3mm}
\caption{$\mu$ bound example
with $\tan\beta\!=\!50$, $m_A\!=\!100$ GeV, $M_{\tilde{Q}}\!=\!400$ GeV,
$M_{\tilde{t}}\!=\!600$ GeV, $M_{\tilde{b}}\!=\!630$ GeV, $A_t\!=\!1$
TeV, and $A_b\!=\!1.3$ TeV. 135 GeV $\!<\!m_{\sba}\!\!<\!$ 404 GeV
varies with $\mu$. $|x|$ is the maximal eigenvalue of $\ct_{fi}^0$ 
as modified in Eq.~(6).} 
\label{fig:mu}
\end{figure}

The approach of the unitarity bound for large trilinear couplings is
illustrated in the figures. MSSM parameters
for these two examples are given in the captions.
Figure \ref{fig:mu} strongly constraints $\mu$, because $A_b$ and $A_t$
have been set only slightly below the bound from Eq. \eqref{eq:Avac} and a high
$\tan\beta$ has been chosen. 
Alternatively, one could consider $A_b$ bounds for large, fixed $\mu$ in
this scenario. In Figure~\ref{fig:mu} the diagonalized amplitude mainly 
results from the interfering low energy processes
$\sba\asba\!\leftrightarrow\!h^0h^0$, 
$\sba\asba\!\leftrightarrow\!h^0H^0$,
$\sba\asba\!\leftrightarrow\!\sba\asba$, and
$\sba\asba\!\leftrightarrow\!H^0H^0$. 
 
\begin{figure}
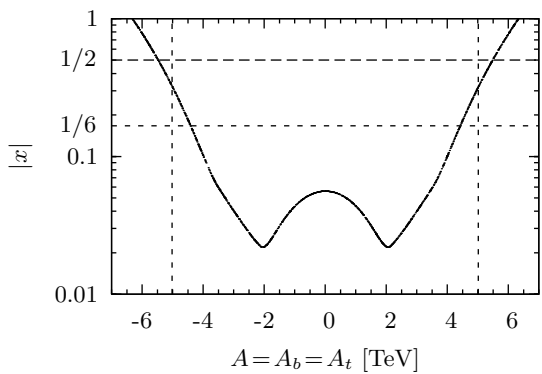

\include{schuessler_susy07fig3} 
\vspace{-3.45mm}
\caption{$A$ bound example, together with the bound from Eq. \eqref{eq:Atvac} (vertical lines) using $\tan\beta\!=\!30$, $m_A\!=\!1$ TeV, $\mu\!=\!-200$ GeV, $M_{\tilde{Q}}\!=\!M_{\tilde{b}}\!=\!2.9$ TeV
with fixed $m_{\sta}\!=\!100$ GeV and $m_{h^0}\!=\!120$ GeV.}
\label{fig:A}
\end{figure}

Figure \ref{fig:A} shows an example for the case of a light $\sta$ held
at a fixed mass of 100 GeV by adjusting $M_{\tilde{t}}^2$ for each value
of $\mu$. To evade the bounds \eqref{eq:Avac} for $A\!=\!A_b\!=\!A_t$,
$M_{\tilde{Q}}$ is chosen high, which leads to the maximum mixing case,
where the $\sta,\asta$ couplings to the Higgs bosons are strongly reduced. 
The largest amplitude is almost completely due to the
$\sta\asta\leftrightarrow h^0h^0$ $t$-channel $\sta$ exchange diagram
and is mostly independent of $\tan\beta\!\gg\!1$, $m_A\!\gg\!m_Z$,
$|\mu|\!\ll\!M_{\tilde{Q}}$ and $M_{\tilde{b}}$. In this scenario
Eq. \eqref{eq:Atvac} gives $|A|\!\lesssim \!5$ TeV, while the
perturbativity condition $|x|\!\leq\!1/6$
yields $|A|\!\lesssim\!4.4$ TeV. 
\section{Conclusion} 
\label{sec:observations}
A general method has been developed for constraining trilinear couplings 
of scalars by using perturbative unitarity. It has been implemented for 
the MSSM, more specifically for third generation squarks and Higgs bos\-ons, 
but it can also be used for other models.
Since trilinear couplings of scalars correspond to superrenormalizable
dimension three operators, they produce the largest contributions to
scattering amplitudes at modest center-of-mass energy. Thus,
perturbative unitarity bounds are derived from scattering amplitudes at
energies not very far above production threshold and they are therefore
independent of the ultraviolet structure of the theory. One finds that
trilinear couplings cannot be much larger than the masses of the scalar
particles involved. Quantitatively, ratios of more than about a factor
5 are forbidden (see Eq.~\eqref{eq:toy}\,). 

We have analyzed 
two MSSM scenarios with small
sbottom or stop masses, trying to constrain the soft SUSY breaking 
parameters $A_b$ and $A_t$ and the Higgs mixing parameter $\mu$ at large
$\tan\beta$. Mixing effects between left- and right-handed squarks lead
to a complicated picture, since large $A$ parameters may lead to only
modest trilinear couplings of the lightest scalars when the relevant
mixing angle factors are small. We have built a numerical program to
study these effects for two Higgs doublets and 
third generation squarks within the MSSM for 
arbitrary input parameters $\tan\beta$, $\mu$, $m_A$, $A_b$, $A_t$,
$M_{\tilde{Q}}^2$, $M_{\tilde{b}}^2$, and $M_{\tilde{t}}^2$. 
For small $\mu$, resulting constraints are similar in strength to bounds
derived from the exclusion of false vacua e.g. Eq.~\eqref{eq:Avac}, 
however, the underlying arguments
are quite different. Our approach therefore provides a complementary 
method for constraining soft SUSY breaking parameters. Given a specific 
MSSM scenario, specified by the Lagrangian parameters,
our program gives a test whether these parameters clash with unitarity.

Since supersymmetric theories are at heart perturbative, even stronger
constraints are obtained by the requirement that scattering amplitudes
sufficiently far from resonances do not leave the perturbative domain.
Such a test is also possible with our program, e.g. by requiring that
the largest eigenvalue of the scattering matrix stays well below the
unitarity bound of $1/2$. Such an analysis becomes instructive when 
analyzing higher loop effects involving SUSY particles which potentially
lead to large corrections of parameters for SM particles, like
Higgs-fermion Yukawa couplings or the mass of the lightest 
$\mathit{CP}$-even Higgs
boson. We have not yet performed such a systematic analysis.

\end{document}

%% file: schuessler_susy07fig2.tex
\begingroup%
  \makeatletter%
  \newcommand{\GNUPLOTspecial}{%
    \@sanitize\catcode`\%=14\relax\special}%
  \setlength{\unitlength}{0.1bp}%
\begin{picture}(2160,1425)(0,0)%
{\GNUPLOTspecial{"
/gnudict 256 dict def
gnudict begin
/Color false def
/Solid false def
/gnulinewidth 5.000 def
/userlinewidth gnulinewidth def
/vshift -33 def
/dl {10.0 mul} def
/hpt_ 31.5 def
/vpt_ 31.5 def
/hpt hpt_ def
/vpt vpt_ def
/Rounded false def
/M {moveto} bind def
/L {lineto} bind def
/R {rmoveto} bind def
/V {rlineto} bind def
/N {newpath moveto} bind def
/C {setrgbcolor} bind def
/f {rlineto fill} bind def
/vpt2 vpt 2 mul def
/hpt2 hpt 2 mul def
/Lshow { currentpoint stroke M
  0 vshift R show } def
/Rshow { currentpoint stroke M
  dup stringwidth pop neg vshift R show } def
/Cshow { currentpoint stroke M
  dup stringwidth pop -2 div vshift R show } def
/UP { dup vpt_ mul /vpt exch def hpt_ mul /hpt exch def
  /hpt2 hpt 2 mul def /vpt2 vpt 2 mul def } def
/DL { Color {setrgbcolor Solid {pop []} if 0 setdash }
 {pop pop pop 0 setgray Solid {pop []} if 0 setdash} ifelse } def
/BL { stroke userlinewidth 2 mul setlinewidth
      Rounded { 1 setlinejoin 1 setlinecap } if } def
/AL { stroke userlinewidth 2 div setlinewidth
      Rounded { 1 setlinejoin 1 setlinecap } if } def
/UL { dup gnulinewidth mul /userlinewidth exch def
      dup 1 lt {pop 1} if 10 mul /udl exch def } def
/PL { stroke userlinewidth setlinewidth
      Rounded { 1 setlinejoin 1 setlinecap } if } def
/LTw { PL [] 1 setgray } def
/LTb { BL [] 0 0 0 DL } def
/LTa { AL [1 udl mul 2 udl mul] 0 setdash 0 0 0 setrgbcolor } def
/LT0 { PL [] 1 0 0 DL } def
/LT1 { PL [4 dl 2 dl] 0 1 0 DL } def
/LT2 { PL [2 dl 3 dl] 0 0 1 DL } def
/LT3 { PL [1 dl 1.5 dl] 1 0 1 DL } def
/LT4 { PL [5 dl 2 dl 1 dl 2 dl] 0 1 1 DL } def
/LT5 { PL [4 dl 3 dl 1 dl 3 dl] 1 1 0 DL } def
/LT6 { PL [2 dl 2 dl 2 dl 4 dl] 0 0 0 DL } def
/LT7 { PL [2 dl 2 dl 2 dl 2 dl 2 dl 4 dl] 1 0.3 0 DL } def
/LT8 { PL [2 dl 2 dl 2 dl 2 dl 2 dl 2 dl 2 dl 4 dl] 0.5 0.5 0.5 DL } def
/Pnt { stroke [] 0 setdash
   gsave 1 setlinecap M 0 0 V stroke grestore } def
/Dia { stroke [] 0 setdash 2 copy vpt add M
  hpt neg vpt neg V hpt vpt neg V
  hpt vpt V hpt neg vpt V closepath stroke
  Pnt } def
/Pls { stroke [] 0 setdash vpt sub M 0 vpt2 V
  currentpoint stroke M
  hpt neg vpt neg R hpt2 0 V stroke
  } def
/Box { stroke [] 0 setdash 2 copy exch hpt sub exch vpt add M
  0 vpt2 neg V hpt2 0 V 0 vpt2 V
  hpt2 neg 0 V closepath stroke
  Pnt } def
/Crs { stroke [] 0 setdash exch hpt sub exch vpt add M
  hpt2 vpt2 neg V currentpoint stroke M
  hpt2 neg 0 R hpt2 vpt2 V stroke } def
/TriU { stroke [] 0 setdash 2 copy vpt 1.12 mul add M
  hpt neg vpt -1.62 mul V
  hpt 2 mul 0 V
  hpt neg vpt 1.62 mul V closepath stroke
  Pnt  } def
/Star { 2 copy Pls Crs } def
/BoxF { stroke [] 0 setdash exch hpt sub exch vpt add M
  0 vpt2 neg V  hpt2 0 V  0 vpt2 V
  hpt2 neg 0 V  closepath fill } def
/TriUF { stroke [] 0 setdash vpt 1.12 mul add M
  hpt neg vpt -1.62 mul V
  hpt 2 mul 0 V
  hpt neg vpt 1.62 mul V closepath fill } def
/TriD { stroke [] 0 setdash 2 copy vpt 1.12 mul sub M
  hpt neg vpt 1.62 mul V
  hpt 2 mul 0 V
  hpt neg vpt -1.62 mul V closepath stroke
  Pnt  } def
/TriDF { stroke [] 0 setdash vpt 1.12 mul sub M
  hpt neg vpt 1.62 mul V
  hpt 2 mul 0 V
  hpt neg vpt -1.62 mul V closepath fill} def
/DiaF { stroke [] 0 setdash vpt add M
  hpt neg vpt neg V hpt vpt neg V
  hpt vpt V hpt neg vpt V closepath fill } def
/Pent { stroke [] 0 setdash 2 copy gsave
  translate 0 hpt M 4 {72 rotate 0 hpt L} repeat
  closepath stroke grestore Pnt } def
/PentF { stroke [] 0 setdash gsave
  translate 0 hpt M 4 {72 rotate 0 hpt L} repeat
  closepath fill grestore } def
/Circle { stroke [] 0 setdash 2 copy
  hpt 0 360 arc stroke Pnt } def
/CircleF { stroke [] 0 setdash hpt 0 360 arc fill } def
/C0 { BL [] 0 setdash 2 copy moveto vpt 90 450  arc } bind def
/C1 { BL [] 0 setdash 2 copy        moveto
       2 copy  vpt 0 90 arc closepath fill
               vpt 0 360 arc closepath } bind def
/C2 { BL [] 0 setdash 2 copy moveto
       2 copy  vpt 90 180 arc closepath fill
               vpt 0 360 arc closepath } bind def
/C3 { BL [] 0 setdash 2 copy moveto
       2 copy  vpt 0 180 arc closepath fill
               vpt 0 360 arc closepath } bind def
/C4 { BL [] 0 setdash 2 copy moveto
       2 copy  vpt 180 270 arc closepath fill
               vpt 0 360 arc closepath } bind def
/C5 { BL [] 0 setdash 2 copy moveto
       2 copy  vpt 0 90 arc
       2 copy moveto
       2 copy  vpt 180 270 arc closepath fill
               vpt 0 360 arc } bind def
/C6 { BL [] 0 setdash 2 copy moveto
      2 copy  vpt 90 270 arc closepath fill
              vpt 0 360 arc closepath } bind def
/C7 { BL [] 0 setdash 2 copy moveto
      2 copy  vpt 0 270 arc closepath fill
              vpt 0 360 arc closepath } bind def
/C8 { BL [] 0 setdash 2 copy moveto
      2 copy vpt 270 360 arc closepath fill
              vpt 0 360 arc closepath } bind def
/C9 { BL [] 0 setdash 2 copy moveto
      2 copy  vpt 270 450 arc closepath fill
              vpt 0 360 arc closepath } bind def
/C10 { BL [] 0 setdash 2 copy 2 copy moveto vpt 270 360 arc closepath fill
       2 copy moveto
       2 copy vpt 90 180 arc closepath fill
               vpt 0 360 arc closepath } bind def
/C11 { BL [] 0 setdash 2 copy moveto
       2 copy  vpt 0 180 arc closepath fill
       2 copy moveto
       2 copy  vpt 270 360 arc closepath fill
               vpt 0 360 arc closepath } bind def
/C12 { BL [] 0 setdash 2 copy moveto
       2 copy  vpt 180 360 arc closepath fill
               vpt 0 360 arc closepath } bind def
/C13 { BL [] 0 setdash  2 copy moveto
       2 copy  vpt 0 90 arc closepath fill
       2 copy moveto
       2 copy  vpt 180 360 arc closepath fill
               vpt 0 360 arc closepath } bind def
/C14 { BL [] 0 setdash 2 copy moveto
       2 copy  vpt 90 360 arc closepath fill
               vpt 0 360 arc } bind def
/C15 { BL [] 0 setdash 2 copy vpt 0 360 arc closepath fill
               vpt 0 360 arc closepath } bind def
/Rec   { newpath 4 2 roll moveto 1 index 0 rlineto 0 exch rlineto
       neg 0 rlineto closepath } bind def
/Square { dup Rec } bind def
/Bsquare { vpt sub exch vpt sub exch vpt2 Square } bind def
/S0 { BL [] 0 setdash 2 copy moveto 0 vpt rlineto BL Bsquare } bind def
/S1 { BL [] 0 setdash 2 copy vpt Square fill Bsquare } bind def
/S2 { BL [] 0 setdash 2 copy exch vpt sub exch vpt Square fill Bsquare } bind def
/S3 { BL [] 0 setdash 2 copy exch vpt sub exch vpt2 vpt Rec fill Bsquare } bind def
/S4 { BL [] 0 setdash 2 copy exch vpt sub exch vpt sub vpt Square fill Bsquare } bind def
/S5 { BL [] 0 setdash 2 copy 2 copy vpt Square fill
       exch vpt sub exch vpt sub vpt Square fill Bsquare } bind def
/S6 { BL [] 0 setdash 2 copy exch vpt sub exch vpt sub vpt vpt2 Rec fill Bsquare } bind def
/S7 { BL [] 0 setdash 2 copy exch vpt sub exch vpt sub vpt vpt2 Rec fill
       2 copy vpt Square fill
       Bsquare } bind def
/S8 { BL [] 0 setdash 2 copy vpt sub vpt Square fill Bsquare } bind def
/S9 { BL [] 0 setdash 2 copy vpt sub vpt vpt2 Rec fill Bsquare } bind def
/S10 { BL [] 0 setdash 2 copy vpt sub vpt Square fill 2 copy exch vpt sub exch vpt Square fill
       Bsquare } bind def
/S11 { BL [] 0 setdash 2 copy vpt sub vpt Square fill 2 copy exch vpt sub exch vpt2 vpt Rec fill
       Bsquare } bind def
/S12 { BL [] 0 setdash 2 copy exch vpt sub exch vpt sub vpt2 vpt Rec fill Bsquare } bind def
/S13 { BL [] 0 setdash 2 copy exch vpt sub exch vpt sub vpt2 vpt Rec fill
       2 copy vpt Square fill Bsquare } bind def
/S14 { BL [] 0 setdash 2 copy exch vpt sub exch vpt sub vpt2 vpt Rec fill
       2 copy exch vpt sub exch vpt Square fill Bsquare } bind def
/S15 { BL [] 0 setdash 2 copy Bsquare fill Bsquare } bind def
/D0 { gsave translate 45 rotate 0 0 S0 stroke grestore } bind def
/D1 { gsave translate 45 rotate 0 0 S1 stroke grestore } bind def
/D2 { gsave translate 45 rotate 0 0 S2 stroke grestore } bind def
/D3 { gsave translate 45 rotate 0 0 S3 stroke grestore } bind def
/D4 { gsave translate 45 rotate 0 0 S4 stroke grestore } bind def
/D5 { gsave translate 45 rotate 0 0 S5 stroke grestore } bind def
/D6 { gsave translate 45 rotate 0 0 S6 stroke grestore } bind def
/D7 { gsave translate 45 rotate 0 0 S7 stroke grestore } bind def
/D8 { gsave translate 45 rotate 0 0 S8 stroke grestore } bind def
/D9 { gsave translate 45 rotate 0 0 S9 stroke grestore } bind def
/D10 { gsave translate 45 rotate 0 0 S10 stroke grestore } bind def
/D11 { gsave translate 45 rotate 0 0 S11 stroke grestore } bind def
/D12 { gsave translate 45 rotate 0 0 S12 stroke grestore } bind def
/D13 { gsave translate 45 rotate 0 0 S13 stroke grestore } bind def
/D14 { gsave translate 45 rotate 0 0 S14 stroke grestore } bind def
/D15 { gsave translate 45 rotate 0 0 S15 stroke grestore } bind def
/DiaE { stroke [] 0 setdash vpt add M
  hpt neg vpt neg V hpt vpt neg V
  hpt vpt V hpt neg vpt V closepath stroke } def
/BoxE { stroke [] 0 setdash exch hpt sub exch vpt add M
  0 vpt2 neg V hpt2 0 V 0 vpt2 V
  hpt2 neg 0 V closepath stroke } def
/TriUE { stroke [] 0 setdash vpt 1.12 mul add M
  hpt neg vpt -1.62 mul V
  hpt 2 mul 0 V
  hpt neg vpt 1.62 mul V closepath stroke } def
/TriDE { stroke [] 0 setdash vpt 1.12 mul sub M
  hpt neg vpt 1.62 mul V
  hpt 2 mul 0 V
  hpt neg vpt -1.62 mul V closepath stroke } def
/PentE { stroke [] 0 setdash gsave
  translate 0 hpt M 4 {72 rotate 0 hpt L} repeat
  closepath stroke grestore } def
/CircE { stroke [] 0 setdash 
  hpt 0 360 arc stroke } def
/Opaque { gsave closepath 1 setgray fill grestore 0 setgray closepath } def
/DiaW { stroke [] 0 setdash vpt add M
  hpt neg vpt neg V hpt vpt neg V
  hpt vpt V hpt neg vpt V Opaque stroke } def
/BoxW { stroke [] 0 setdash exch hpt sub exch vpt add M
  0 vpt2 neg V hpt2 0 V 0 vpt2 V
  hpt2 neg 0 V Opaque stroke } def
/TriUW { stroke [] 0 setdash vpt 1.12 mul add M
  hpt neg vpt -1.62 mul V
  hpt 2 mul 0 V
  hpt neg vpt 1.62 mul V Opaque stroke } def
/TriDW { stroke [] 0 setdash vpt 1.12 mul sub M
  hpt neg vpt 1.62 mul V
  hpt 2 mul 0 V
  hpt neg vpt -1.62 mul V Opaque stroke } def
/PentW { stroke [] 0 setdash gsave
  translate 0 hpt M 4 {72 rotate 0 hpt L} repeat
  Opaque stroke grestore } def
/CircW { stroke [] 0 setdash 
  hpt 0 360 arc Opaque stroke } def
/BoxFill { gsave Rec 1 setgray fill grestore } def
/BoxColFill {
  gsave Rec
  /Fillden exch def
  currentrgbcolor
  /ColB exch def /ColG exch def /ColR exch def
  /ColR ColR Fillden mul Fillden sub 1 add def
  /ColG ColG Fillden mul Fillden sub 1 add def
  /ColB ColB Fillden mul Fillden sub 1 add def
  ColR ColG ColB setrgbcolor
  fill grestore } def
%
%
/PatternFill { gsave /PFa [ 9 2 roll ] def
    PFa 0 get PFa 2 get 2 div add PFa 1 get PFa 3 get 2 div add translate
    PFa 2 get -2 div PFa 3 get -2 div PFa 2 get PFa 3 get Rec
    gsave 1 setgray fill grestore clip
    currentlinewidth 0.5 mul setlinewidth
    /PFs PFa 2 get dup mul PFa 3 get dup mul add sqrt def
    0 0 M PFa 5 get rotate PFs -2 div dup translate
	0 1 PFs PFa 4 get div 1 add floor cvi
	{ PFa 4 get mul 0 M 0 PFs V } for
    0 PFa 6 get ne {
	0 1 PFs PFa 4 get div 1 add floor cvi
	{ PFa 4 get mul 0 2 1 roll M PFs 0 V } for
    } if
    stroke grestore } def
/Symbol-Oblique /Symbol findfont [1 0 .167 1 0 0] makefont
dup length dict begin {1 index /FID eq {pop pop} {def} ifelse} forall
currentdict end definefont pop
end
gnudict begin
gsave
0 0 translate
0.100 0.100 scale
0 setgray
newpath
0.500 UL
LTb
400 537 M
37 0 V
1573 0 R
-37 0 V
0.500 UL
LTb
400 300 M
18 0 V
1592 0 R
-18 0 V
400 362 M
18 0 V
1592 0 R
-18 0 V
400 415 M
18 0 V
1592 0 R
-18 0 V
400 461 M
18 0 V
1592 0 R
-18 0 V
400 501 M
18 0 V
1592 0 R
-18 0 V
400 537 M
18 0 V
1592 0 R
-18 0 V
400 1326 M
37 0 V
1573 0 R
-37 0 V
0.500 UL
LTb
400 537 M
18 0 V
1592 0 R
-18 0 V
400 775 M
18 0 V
1592 0 R
-18 0 V
400 914 M
18 0 V
1592 0 R
-18 0 V
400 1012 M
18 0 V
1592 0 R
-18 0 V
400 1089 M
18 0 V
1592 0 R
-18 0 V
400 1151 M
18 0 V
1592 0 R
-18 0 V
400 1204 M
18 0 V
1592 0 R
-18 0 V
400 1250 M
18 0 V
1592 0 R
-18 0 V
400 1290 M
18 0 V
1592 0 R
-18 0 V
400 1326 M
18 0 V
1592 0 R
-18 0 V
400 300 M
0 37 V
0 989 R
0 -37 V
0.500 UL
LTb
454 300 M
0 18 V
0 1008 R
0 -18 V
507 300 M
0 18 V
0 1008 R
0 -18 V
561 300 M
0 18 V
0 1008 R
0 -18 V
615 300 M
0 18 V
0 1008 R
0 -18 V
668 300 M
0 37 V
0 989 R
0 -37 V
0.500 UL
LTb
722 300 M
0 18 V
0 1008 R
0 -18 V
776 300 M
0 18 V
0 1008 R
0 -18 V
829 300 M
0 18 V
0 1008 R
0 -18 V
883 300 M
0 18 V
0 1008 R
0 -18 V
937 300 M
0 37 V
0 989 R
0 -37 V
0.500 UL
LTb
990 300 M
0 18 V
0 1008 R
0 -18 V
1044 300 M
0 18 V
0 1008 R
0 -18 V
1098 300 M
0 18 V
0 1008 R
0 -18 V
1151 300 M
0 18 V
0 1008 R
0 -18 V
1205 300 M
0 37 V
0 989 R
0 -37 V
0.500 UL
LTb
1259 300 M
0 18 V
0 1008 R
0 -18 V
1312 300 M
0 18 V
0 1008 R
0 -18 V
1366 300 M
0 18 V
0 1008 R
0 -18 V
1420 300 M
0 18 V
0 1008 R
0 -18 V
1473 300 M
0 37 V
0 989 R
0 -37 V
0.500 UL
LTb
1527 300 M
0 18 V
0 1008 R
0 -18 V
1581 300 M
0 18 V
0 1008 R
0 -18 V
1634 300 M
0 18 V
0 1008 R
0 -18 V
1688 300 M
0 18 V
0 1008 R
0 -18 V
1742 300 M
0 37 V
0 989 R
0 -37 V
0.500 UL
LTb
1795 300 M
0 18 V
0 1008 R
0 -18 V
1849 300 M
0 18 V
0 1008 R
0 -18 V
1903 300 M
0 18 V
0 1008 R
0 -18 V
1956 300 M
0 18 V
0 1008 R
0 -18 V
2010 300 M
0 37 V
0 989 R
0 -37 V
0.500 UL
LTb
0.500 UL
LTb
400 300 M
1610 0 V
0 1026 V
-1610 0 V
400 300 L
LTb
LTb
1.000 UP
LTb
LTb
1.000 UL
LT0
1823 929 Pnt
1730 812 Pnt
1178 394 Pnt
472 1091 Pnt
1974 1239 Pnt
1270 400 Pnt
1286 403 Pnt
709 760 Pnt
734 733 Pnt
521 1002 Pnt
1447 510 Pnt
471 1093 Pnt
886 575 Pnt
1618 687 Pnt
416 1228 Pnt
540 970 Pnt
687 784 Pnt
1072 416 Pnt
1882 1026 Pnt
807 657 Pnt
1992 1298 Pnt
1562 627 Pnt
1198 393 Pnt
1210 393 Pnt
539 972 Pnt
713 756 Pnt
815 648 Pnt
1998 1321 Pnt
1415 480 Pnt
1900 1060 Pnt
993 474 Pnt
1396 464 Pnt
1458 520 Pnt
1952 1177 Pnt
961 502 Pnt
1740 823 Pnt
1050 427 Pnt
1150 397 Pnt
787 678 Pnt
937 525 Pnt
685 787 Pnt
1933 1130 Pnt
1142 398 Pnt
740 728 Pnt
1476 538 Pnt
1501 563 Pnt
599 882 Pnt
1684 760 Pnt
1413 478 Pnt
815 648 Pnt
1897 1055 Pnt
1155 396 Pnt
1534 598 Pnt
947 516 Pnt
1757 843 Pnt
1148 397 Pnt
1237 395 Pnt
1838 953 Pnt
729 739 Pnt
1561 626 Pnt
1857 984 Pnt
1931 1125 Pnt
1650 723 Pnt
1147 397 Pnt
1560 625 Pnt
751 716 Pnt
1268 399 Pnt
1230 394 Pnt
1537 601 Pnt
907 554 Pnt
696 775 Pnt
1886 1033 Pnt
1372 444 Pnt
549 956 Pnt
1129 400 Pnt
1326 415 Pnt
1834 947 Pnt
658 816 Pnt
1906 1072 Pnt
919 543 Pnt
1562 627 Pnt
1067 418 Pnt
469 1097 Pnt
1753 838 Pnt
1490 552 Pnt
1587 654 Pnt
1205 393 Pnt
1056 423 Pnt
1355 430 Pnt
1299 407 Pnt
1101 407 Pnt
1657 730 Pnt
1927 1116 Pnt
841 621 Pnt
683 789 Pnt
780 685 Pnt
982 483 Pnt
601 880 Pnt
1635 706 Pnt
1382 452 Pnt
1899 1058 Pnt
1441 504 Pnt
1157 396 Pnt
975 489 Pnt
853 608 Pnt
1589 656 Pnt
571 923 Pnt
1309 410 Pnt
1490 553 Pnt
1339 420 Pnt
447 1146 Pnt
439 1165 Pnt
1418 483 Pnt
938 524 Pnt
881 581 Pnt
652 822 Pnt
429 1190 Pnt
895 566 Pnt
703 767 Pnt
1072 416 Pnt
1415 480 Pnt
942 520 Pnt
1105 406 Pnt
517 1007 Pnt
1743 827 Pnt
981 484 Pnt
1246 396 Pnt
1669 743 Pnt
473 1090 Pnt
693 777 Pnt
1373 444 Pnt
947 516 Pnt
1186 393 Pnt
1488 550 Pnt
1995 1312 Pnt
1418 483 Pnt
552 951 Pnt
1409 475 Pnt
982 483 Pnt
592 893 Pnt
1287 404 Pnt
840 622 Pnt
631 845 Pnt
1281 402 Pnt
539 972 Pnt
1210 393 Pnt
834 628 Pnt
1334 418 Pnt
1343 421 Pnt
1115 404 Pnt
1306 409 Pnt
688 783 Pnt
1249 396 Pnt
896 565 Pnt
1067 418 Pnt
611 867 Pnt
1692 770 Pnt
1684 761 Pnt
1998 1322 Pnt
1430 494 Pnt
1774 862 Pnt
1889 1040 Pnt
400 1276 Pnt
1598 666 Pnt
789 675 Pnt
1749 834 Pnt
1772 859 Pnt
648 826 Pnt
959 504 Pnt
1166 395 Pnt
1306 409 Pnt
558 942 Pnt
1579 645 Pnt
975 489 Pnt
1795 888 Pnt
650 824 Pnt
1862 992 Pnt
1735 818 Pnt
1628 698 Pnt
1348 425 Pnt
941 521 Pnt
1215 393 Pnt
799 665 Pnt
1319 413 Pnt
522 1000 Pnt
553 950 Pnt
814 649 Pnt
725 743 Pnt
1138 399 Pnt
1918 1097 Pnt
514 1013 Pnt
921 540 Pnt
1827 936 Pnt
1186 393 Pnt
964 499 Pnt
1396 464 Pnt
1516 578 Pnt
1287 404 Pnt
513 1015 Pnt
1413 478 Pnt
1396 463 Pnt
1435 498 Pnt
1113 404 Pnt
1511 574 Pnt
1551 615 Pnt
1423 487 Pnt
1346 424 Pnt
1827 936 Pnt
848 614 Pnt
829 633 Pnt
688 783 Pnt
1216 393 Pnt
1404 470 Pnt
869 592 Pnt
1575 641 Pnt
1005 463 Pnt
911 551 Pnt
499 1039 Pnt
1551 616 Pnt
1898 1055 Pnt
943 519 Pnt
582 907 Pnt
1755 840 Pnt
1692 770 Pnt
592 893 Pnt
1664 739 Pnt
1874 1013 Pnt
621 855 Pnt
408 1252 Pnt
961 502 Pnt
786 679 Pnt
1014 455 Pnt
1584 650 Pnt
1014 455 Pnt
1125 401 Pnt
1367 440 Pnt
1911 1082 Pnt
1920 1101 Pnt
677 795 Pnt
767 698 Pnt
481 1073 Pnt
1769 856 Pnt
1377 447 Pnt
427 1196 Pnt
1162 395 Pnt
1514 577 Pnt
1392 460 Pnt
1838 953 Pnt
1717 797 Pnt
791 673 Pnt
1564 629 Pnt
1440 503 Pnt
1923 1107 Pnt
956 507 Pnt
446 1148 Pnt
1874 1013 Pnt
1581 647 Pnt
1714 794 Pnt
541 968 Pnt
801 662 Pnt
1428 492 Pnt
733 735 Pnt
1508 571 Pnt
1483 545 Pnt
1825 933 Pnt
910 551 Pnt
1449 512 Pnt
527 991 Pnt
1137 399 Pnt
1475 538 Pnt
568 927 Pnt
1761 847 Pnt
418 1221 Pnt
1420 485 Pnt
1771 858 Pnt
697 774 Pnt
1374 445 Pnt
564 933 Pnt
1880 1024 Pnt
1908 1077 Pnt
983 482 Pnt
1504 566 Pnt
1284 403 Pnt
730 738 Pnt
1089 411 Pnt
680 793 Pnt
1216 393 Pnt
655 819 Pnt
866 595 Pnt
1452 515 Pnt
1193 393 Pnt
941 521 Pnt
845 617 Pnt
694 777 Pnt
984 481 Pnt
545 962 Pnt
974 490 Pnt
1133 400 Pnt
511 1018 Pnt
1528 591 Pnt
726 742 Pnt
1975 1241 Pnt
935 527 Pnt
1237 395 Pnt
824 638 Pnt
1719 800 Pnt
1629 699 Pnt
1068 418 Pnt
1230 394 Pnt
1700 779 Pnt
1063 419 Pnt
1306 409 Pnt
1447 510 Pnt
1305 408 Pnt
1969 1223 Pnt
1397 464 Pnt
1585 651 Pnt
1748 832 Pnt
926 535 Pnt
1238 395 Pnt
1311 410 Pnt
1884 1031 Pnt
511 1018 Pnt
1737 820 Pnt
1960 1198 Pnt
1082 413 Pnt
1071 417 Pnt
1618 688 Pnt
1608 676 Pnt
822 640 Pnt
1192 393 Pnt
1847 967 Pnt
536 977 Pnt
1557 622 Pnt
1388 457 Pnt
852 609 Pnt
500 1038 Pnt
1188 393 Pnt
1754 840 Pnt
1899 1058 Pnt
1192 393 Pnt
869 593 Pnt
1800 895 Pnt
1678 754 Pnt
655 819 Pnt
1020 450 Pnt
589 897 Pnt
1605 674 Pnt
709 761 Pnt
888 573 Pnt
988 477 Pnt
1287 404 Pnt
773 692 Pnt
1976 1244 Pnt
1500 562 Pnt
1519 582 Pnt
997 470 Pnt
1605 673 Pnt
1993 1303 Pnt
660 813 Pnt
532 984 Pnt
564 934 Pnt
514 1013 Pnt
888 573 Pnt
1219 394 Pnt
953 509 Pnt
1795 889 Pnt
1373 444 Pnt
1565 630 Pnt
487 1061 Pnt
1528 592 Pnt
703 767 Pnt
1817 920 Pnt
997 470 Pnt
1520 583 Pnt
1071 417 Pnt
535 979 Pnt
776 689 Pnt
754 712 Pnt
1863 993 Pnt
1088 411 Pnt
1402 469 Pnt
847 615 Pnt
1406 472 Pnt
474 1088 Pnt
505 1030 Pnt
793 671 Pnt
1166 395 Pnt
1212 393 Pnt
1158 396 Pnt
1178 394 Pnt
1663 737 Pnt
1004 464 Pnt
1468 531 Pnt
526 993 Pnt
1056 423 Pnt
1047 430 Pnt
1295 406 Pnt
500 1038 Pnt
1082 413 Pnt
1418 483 Pnt
1925 1112 Pnt
1851 973 Pnt
1017 453 Pnt
1331 417 Pnt
727 741 Pnt
608 870 Pnt
1964 1208 Pnt
1379 449 Pnt
690 782 Pnt
1988 1283 Pnt
1211 393 Pnt
720 748 Pnt
554 949 Pnt
667 806 Pnt
971 493 Pnt
1461 523 Pnt
1071 417 Pnt
1248 396 Pnt
1386 455 Pnt
1420 485 Pnt
685 786 Pnt
1358 432 Pnt
1339 420 Pnt
484 1069 Pnt
1042 433 Pnt
545 962 Pnt
951 512 Pnt
1193 393 Pnt
1396 464 Pnt
1118 403 Pnt
604 876 Pnt
971 493 Pnt
798 665 Pnt
1000 467 Pnt
530 986 Pnt
893 568 Pnt
1953 1178 Pnt
1555 620 Pnt
1365 438 Pnt
1945 1158 Pnt
1918 1096 Pnt
614 862 Pnt
1946 1162 Pnt
1767 854 Pnt
1383 452 Pnt
1510 573 Pnt
1774 861 Pnt
886 576 Pnt
1112 404 Pnt
501 1037 Pnt
1304 408 Pnt
1846 966 Pnt
1150 397 Pnt
1279 402 Pnt
585 902 Pnt
412 1239 Pnt
577 914 Pnt
1203 393 Pnt
1094 409 Pnt
1400 467 Pnt
732 735 Pnt
789 675 Pnt
423 1207 Pnt
1395 463 Pnt
691 780 Pnt
1446 509 Pnt
804 659 Pnt
543 966 Pnt
574 918 Pnt
1204 393 Pnt
443 1156 Pnt
1896 1052 Pnt
1781 869 Pnt
1814 916 Pnt
1782 870 Pnt
807 657 Pnt
1812 913 Pnt
1777 865 Pnt
1495 557 Pnt
1581 647 Pnt
807 656 Pnt
689 782 Pnt
1509 572 Pnt
1195 393 Pnt
740 727 Pnt
1662 735 Pnt
1649 722 Pnt
730 737 Pnt
662 811 Pnt
1897 1054 Pnt
1806 904 Pnt
1944 1156 Pnt
1939 1143 Pnt
1601 669 Pnt
716 753 Pnt
628 848 Pnt
1988 1283 Pnt
1229 394 Pnt
1502 565 Pnt
1077 415 Pnt
1841 958 Pnt
1214 393 Pnt
910 551 Pnt
991 475 Pnt
1706 785 Pnt
1890 1041 Pnt
904 558 Pnt
1690 768 Pnt
1719 800 Pnt
746 720 Pnt
1025 446 Pnt
1988 1285 Pnt
1918 1096 Pnt
1490 552 Pnt
1322 414 Pnt
1368 440 Pnt
639 836 Pnt
1643 715 Pnt
1452 515 Pnt
515 1011 Pnt
922 540 Pnt
1530 594 Pnt
1061 420 Pnt
1630 700 Pnt
644 831 Pnt
548 958 Pnt
1053 425 Pnt
1875 1014 Pnt
1817 920 Pnt
1731 814 Pnt
1253 397 Pnt
1622 692 Pnt
594 890 Pnt
1135 399 Pnt
985 480 Pnt
1681 758 Pnt
440 1163 Pnt
540 971 Pnt
974 491 Pnt
1604 672 Pnt
579 912 Pnt
1328 416 Pnt
1214 393 Pnt
1966 1215 Pnt
584 904 Pnt
752 715 Pnt
1815 918 Pnt
1017 453 Pnt
1462 524 Pnt
1215 393 Pnt
1710 789 Pnt
1070 417 Pnt
1434 498 Pnt
1676 752 Pnt
744 723 Pnt
1246 396 Pnt
1239 395 Pnt
430 1188 Pnt
1720 801 Pnt
1392 460 Pnt
1870 1005 Pnt
1311 410 Pnt
1273 400 Pnt
1697 775 Pnt
1410 476 Pnt
1857 983 Pnt
925 536 Pnt
1041 434 Pnt
1434 497 Pnt
650 825 Pnt
1615 684 Pnt
418 1219 Pnt
1181 394 Pnt
1943 1154 Pnt
1237 395 Pnt
1600 668 Pnt
1489 551 Pnt
1815 918 Pnt
521 1002 Pnt
1557 622 Pnt
1664 739 Pnt
944 519 Pnt
1653 726 Pnt
1897 1055 Pnt
645 830 Pnt
460 1116 Pnt
801 663 Pnt
1359 433 Pnt
923 538 Pnt
1813 915 Pnt
1463 525 Pnt
1231 394 Pnt
984 481 Pnt
447 1146 Pnt
684 788 Pnt
1962 1202 Pnt
1588 655 Pnt
1957 1188 Pnt
1198 393 Pnt
1649 722 Pnt
1114 404 Pnt
1978 1249 Pnt
1295 406 Pnt
1364 437 Pnt
1504 567 Pnt
1744 828 Pnt
945 517 Pnt
1775 863 Pnt
448 1143 Pnt
1562 627 Pnt
542 968 Pnt
1927 1117 Pnt
590 896 Pnt
1531 594 Pnt
1118 403 Pnt
1120 402 Pnt
614 862 Pnt
1198 393 Pnt
427 1195 Pnt
782 683 Pnt
849 612 Pnt
1498 561 Pnt
862 599 Pnt
937 525 Pnt
1752 837 Pnt
1914 1088 Pnt
1147 397 Pnt
676 796 Pnt
1569 635 Pnt
1327 416 Pnt
1515 578 Pnt
1256 397 Pnt
734 734 Pnt
1443 506 Pnt
630 845 Pnt
1831 942 Pnt
1423 488 Pnt
1876 1016 Pnt
943 519 Pnt
655 819 Pnt
1783 872 Pnt
1829 939 Pnt
1892 1045 Pnt
1601 669 Pnt
973 491 Pnt
1034 439 Pnt
1405 471 Pnt
435 1175 Pnt
849 613 Pnt
1663 737 Pnt
1423 487 Pnt
1318 412 Pnt
1051 427 Pnt
1862 992 Pnt
783 681 Pnt
1253 397 Pnt
406 1258 Pnt
1002 466 Pnt
1552 617 Pnt
987 479 Pnt
722 746 Pnt
977 487 Pnt
1829 939 Pnt
1858 984 Pnt
1224 394 Pnt
1535 598 Pnt
620 856 Pnt
1339 420 Pnt
1342 421 Pnt
1268 399 Pnt
796 668 Pnt
871 590 Pnt
1026 446 Pnt
1518 581 Pnt
1208 393 Pnt
1633 704 Pnt
1706 785 Pnt
1982 1263 Pnt
1154 396 Pnt
517 1007 Pnt
1765 851 Pnt
449 1141 Pnt
1996 1315 Pnt
1319 413 Pnt
1083 413 Pnt
948 514 Pnt
1230 394 Pnt
1981 1261 Pnt
518 1007 Pnt
1191 393 Pnt
1779 866 Pnt
1981 1261 Pnt
1779 867 Pnt
873 588 Pnt
640 835 Pnt
1734 817 Pnt
1809 909 Pnt
1164 395 Pnt
1268 400 Pnt
1456 519 Pnt
1156 396 Pnt
409 1247 Pnt
736 732 Pnt
716 753 Pnt
923 538 Pnt
999 468 Pnt
1134 399 Pnt
1220 394 Pnt
873 588 Pnt
1171 394 Pnt
1304 408 Pnt
1526 589 Pnt
1155 396 Pnt
1286 403 Pnt
596 888 Pnt
636 839 Pnt
1847 968 Pnt
838 624 Pnt
622 853 Pnt
1868 1001 Pnt
467 1101 Pnt
1647 719 Pnt
1479 541 Pnt
1225 394 Pnt
1613 682 Pnt
1285 403 Pnt
1518 581 Pnt
1436 500 Pnt
1345 423 Pnt
1952 1175 Pnt
1203 393 Pnt
1986 1276 Pnt
1342 421 Pnt
1958 1192 Pnt
1021 449 Pnt
885 576 Pnt
491 1055 Pnt
810 653 Pnt
936 526 Pnt
719 750 Pnt
1343 422 Pnt
869 592 Pnt
1723 804 Pnt
1913 1085 Pnt
1550 615 Pnt
528 989 Pnt
1361 435 Pnt
556 945 Pnt
1129 400 Pnt
859 602 Pnt
1013 456 Pnt
1961 1200 Pnt
1638 709 Pnt
643 832 Pnt
1166 395 Pnt
976 488 Pnt
735 732 Pnt
1856 981 Pnt
1828 938 Pnt
1269 400 Pnt
932 530 Pnt
1722 804 Pnt
1194 393 Pnt
1421 486 Pnt
414 1233 Pnt
1536 600 Pnt
1527 590 Pnt
1425 489 Pnt
1982 1262 Pnt
1549 613 Pnt
1682 758 Pnt
719 749 Pnt
1452 514 Pnt
567 928 Pnt
1177 394 Pnt
693 778 Pnt
1582 648 Pnt
1257 398 Pnt
1993 1304 Pnt
555 947 Pnt
439 1164 Pnt
1501 563 Pnt
1735 818 Pnt
1619 689 Pnt
1132 400 Pnt
1972 1231 Pnt
1221 394 Pnt
1417 482 Pnt
806 657 Pnt
574 918 Pnt
1146 397 Pnt
1529 592 Pnt
1408 474 Pnt
499 1040 Pnt
796 668 Pnt
806 658 Pnt
519 1005 Pnt
976 489 Pnt
797 666 Pnt
1565 630 Pnt
999 468 Pnt
1917 1094 Pnt
1915 1090 Pnt
853 608 Pnt
1455 518 Pnt
921 540 Pnt
800 664 Pnt
1913 1086 Pnt
848 614 Pnt
1284 403 Pnt
603 877 Pnt
1399 466 Pnt
820 643 Pnt
1679 755 Pnt
1606 675 Pnt
1758 844 Pnt
1006 462 Pnt
701 769 Pnt
1721 803 Pnt
584 904 Pnt
1243 396 Pnt
556 945 Pnt
1469 531 Pnt
940 522 Pnt
1708 787 Pnt
1824 930 Pnt
927 534 Pnt
1968 1220 Pnt
982 483 Pnt
1292 405 Pnt
902 559 Pnt
521 1001 Pnt
1186 393 Pnt
1356 431 Pnt
1001 467 Pnt
826 637 Pnt
870 592 Pnt
868 594 Pnt
1237 395 Pnt
880 581 Pnt
605 875 Pnt
628 847 Pnt
614 862 Pnt
1553 618 Pnt
979 486 Pnt
1047 429 Pnt
436 1173 Pnt
1588 655 Pnt
1578 644 Pnt
1422 486 Pnt
908 553 Pnt
571 923 Pnt
786 678 Pnt
1606 675 Pnt
602 878 Pnt
1245 396 Pnt
901 560 Pnt
1337 419 Pnt
734 734 Pnt
1648 720 Pnt
436 1173 Pnt
1192 393 Pnt
1085 412 Pnt
1725 806 Pnt
1139 399 Pnt
499 1040 Pnt
1386 455 Pnt
1096 408 Pnt
1328 416 Pnt
679 793 Pnt
1267 399 Pnt
1521 584 Pnt
1099 408 Pnt
1877 1018 Pnt
470 1095 Pnt
1930 1124 Pnt
1790 881 Pnt
605 875 Pnt
947 515 Pnt
1600 668 Pnt
930 531 Pnt
1119 403 Pnt
1079 414 Pnt
505 1029 Pnt
1308 409 Pnt
1464 526 Pnt
530 986 Pnt
954 509 Pnt
1695 773 Pnt
1378 448 Pnt
1843 961 Pnt
1232 394 Pnt
764 702 Pnt
753 713 Pnt
542 967 Pnt
481 1074 Pnt
1044 431 Pnt
1122 402 Pnt
1299 407 Pnt
937 525 Pnt
1839 954 Pnt
621 855 Pnt
1161 395 Pnt
1216 393 Pnt
1945 1158 Pnt
1720 801 Pnt
1858 985 Pnt
1155 396 Pnt
479 1077 Pnt
1608 676 Pnt
1989 1289 Pnt
899 562 Pnt
1820 925 Pnt
468 1100 Pnt
1652 724 Pnt
1661 734 Pnt
1719 800 Pnt
1800 895 Pnt
1524 587 Pnt
586 900 Pnt
1717 798 Pnt
1663 737 Pnt
1139 399 Pnt
1400 467 Pnt
1776 864 Pnt
1319 413 Pnt
1850 973 Pnt
1945 1159 Pnt
1321 413 Pnt
1204 393 Pnt
1459 521 Pnt
1573 639 Pnt
1628 698 Pnt
1711 791 Pnt
828 635 Pnt
1519 582 Pnt
1661 735 Pnt
1844 962 Pnt
1849 970 Pnt
1940 1145 Pnt
1001 466 Pnt
703 767 Pnt
586 901 Pnt
1930 1123 Pnt
890 571 Pnt
986 480 Pnt
919 542 Pnt
564 934 Pnt
1803 899 Pnt
1519 582 Pnt
510 1019 Pnt
896 565 Pnt
1382 452 Pnt
1785 874 Pnt
917 545 Pnt
928 533 Pnt
574 919 Pnt
862 600 Pnt
1436 500 Pnt
756 710 Pnt
1968 1221 Pnt
493 1051 Pnt
1177 394 Pnt
1473 535 Pnt
1374 445 Pnt
1580 646 Pnt
1192 393 Pnt
782 683 Pnt
905 556 Pnt
712 757 Pnt
1914 1087 Pnt
402 1271 Pnt
1192 393 Pnt
788 676 Pnt
1986 1279 Pnt
1256 397 Pnt
628 848 Pnt
595 888 Pnt
1925 1112 Pnt
1161 395 Pnt
777 687 Pnt
1687 763 Pnt
1285 403 Pnt
615 860 Pnt
1816 920 Pnt
1613 682 Pnt
1317 412 Pnt
1404 471 Pnt
994 473 Pnt
614 862 Pnt
1679 755 Pnt
818 645 Pnt
1981 1260 Pnt
1900 1060 Pnt
857 605 Pnt
564 933 Pnt
1701 780 Pnt
1602 670 Pnt
1600 667 Pnt
1779 866 Pnt
746 721 Pnt
460 1116 Pnt
1559 624 Pnt
1059 421 Pnt
823 640 Pnt
716 753 Pnt
1669 744 Pnt
1227 394 Pnt
879 582 Pnt
1170 394 Pnt
680 792 Pnt
839 623 Pnt
1882 1026 Pnt
1821 927 Pnt
1664 738 Pnt
1889 1039 Pnt
620 856 Pnt
1807 906 Pnt
488 1060 Pnt
1136 399 Pnt
1496 558 Pnt
459 1118 Pnt
894 567 Pnt
670 803 Pnt
523 999 Pnt
468 1101 Pnt
984 481 Pnt
1941 1147 Pnt
1522 585 Pnt
666 807 Pnt
541 969 Pnt
1587 654 Pnt
1926 1113 Pnt
793 671 Pnt
1568 633 Pnt
1369 441 Pnt
1165 395 Pnt
1222 394 Pnt
787 677 Pnt
1589 656 Pnt
1996 1314 Pnt
1138 399 Pnt
1885 1031 Pnt
1177 394 Pnt
1282 402 Pnt
1397 465 Pnt
1787 877 Pnt
437 1170 Pnt
1306 409 Pnt
707 762 Pnt
557 944 Pnt
1586 652 Pnt
477 1080 Pnt
1426 490 Pnt
1463 526 Pnt
1391 459 Pnt
708 762 Pnt
1223 394 Pnt
726 742 Pnt
1372 444 Pnt
536 976 Pnt
1036 438 Pnt
743 724 Pnt
716 753 Pnt
1447 510 Pnt
1759 845 Pnt
1152 397 Pnt
651 824 Pnt
1571 637 Pnt
1029 443 Pnt
1703 782 Pnt
1302 408 Pnt
1748 832 Pnt
1302 408 Pnt
1313 411 Pnt
652 823 Pnt
497 1042 Pnt
718 751 Pnt
1527 591 Pnt
1625 695 Pnt
638 837 Pnt
1517 579 Pnt
1460 522 Pnt
1235 395 Pnt
1599 667 Pnt
1482 544 Pnt
1347 424 Pnt
1490 552 Pnt
1450 513 Pnt
1768 855 Pnt
1513 576 Pnt
1830 940 Pnt
413 1236 Pnt
1205 393 Pnt
1259 398 Pnt
998 469 Pnt
490 1055 Pnt
1246 396 Pnt
1909 1079 Pnt
451 1138 Pnt
1772 859 Pnt
1232 394 Pnt
1664 739 Pnt
622 853 Pnt
805 659 Pnt
1430 494 Pnt
818 645 Pnt
491 1055 Pnt
662 811 Pnt
633 842 Pnt
620 856 Pnt
1280 402 Pnt
654 820 Pnt
1489 551 Pnt
778 687 Pnt
618 858 Pnt
1111 405 Pnt
1008 461 Pnt
1041 434 Pnt
601 880 Pnt
1983 1267 Pnt
533 982 Pnt
1927 1115 Pnt
638 837 Pnt
1558 623 Pnt
1621 691 Pnt
538 974 Pnt
995 472 Pnt
1819 924 Pnt
450 1139 Pnt
962 501 Pnt
553 950 Pnt
1809 909 Pnt
1269 400 Pnt
1012 457 Pnt
688 784 Pnt
1858 985 Pnt
1454 517 Pnt
1329 416 Pnt
805 658 Pnt
1783 872 Pnt
1699 778 Pnt
486 1063 Pnt
909 552 Pnt
1380 450 Pnt
1107 406 Pnt
1171 394 Pnt
784 680 Pnt
1277 401 Pnt
1904 1068 Pnt
510 1020 Pnt
1456 519 Pnt
440 1163 Pnt
1497 560 Pnt
742 725 Pnt
1074 416 Pnt
416 1225 Pnt
1932 1128 Pnt
1176 394 Pnt
942 520 Pnt
1073 416 Pnt
1832 943 Pnt
1385 454 Pnt
1018 452 Pnt
1946 1161 Pnt
1794 887 Pnt
1697 775 Pnt
1431 495 Pnt
1326 415 Pnt
1916 1093 Pnt
1996 1314 Pnt
1347 424 Pnt
1019 451 Pnt
475 1085 Pnt
1810 910 Pnt
1420 485 Pnt
928 533 Pnt
1112 404 Pnt
1946 1160 Pnt
523 998 Pnt
453 1132 Pnt
1227 394 Pnt
1465 528 Pnt
1929 1121 Pnt
1036 438 Pnt
873 589 Pnt
1395 462 Pnt
1987 1281 Pnt
831 631 Pnt
1214 393 Pnt
1404 471 Pnt
824 638 Pnt
597 886 Pnt
864 597 Pnt
1310 410 Pnt
659 814 Pnt
1182 394 Pnt
1456 518 Pnt
673 799 Pnt
815 648 Pnt
1072 416 Pnt
901 561 Pnt
819 644 Pnt
1565 630 Pnt
1775 862 Pnt
438 1168 Pnt
826 637 Pnt
1662 736 Pnt
955 508 Pnt
532 982 Pnt
766 699 Pnt
992 474 Pnt
1508 571 Pnt
1896 1053 Pnt
1398 465 Pnt
1398 465 Pnt
839 623 Pnt
584 905 Pnt
1814 916 Pnt
1725 807 Pnt
1158 396 Pnt
568 927 Pnt
720 749 Pnt
1107 405 Pnt
1451 514 Pnt
1349 426 Pnt
731 737 Pnt
769 697 Pnt
1831 942 Pnt
1916 1093 Pnt
1500 563 Pnt
1683 759 Pnt
928 534 Pnt
693 778 Pnt
1976 1243 Pnt
807 656 Pnt
937 525 Pnt
881 580 Pnt
549 956 Pnt
1276 401 Pnt
981 484 Pnt
1206 393 Pnt
717 752 Pnt
1044 432 Pnt
1757 843 Pnt
1485 548 Pnt
826 637 Pnt
1248 396 Pnt
1186 393 Pnt
1673 748 Pnt
1765 852 Pnt
474 1087 Pnt
1286 404 Pnt
1227 394 Pnt
1890 1041 Pnt
736 732 Pnt
1215 393 Pnt
1949 1169 Pnt
1204 393 Pnt
1192 393 Pnt
1744 828 Pnt
1020 451 Pnt
1059 421 Pnt
1513 575 Pnt
1946 1161 Pnt
666 807 Pnt
827 636 Pnt
965 499 Pnt
880 581 Pnt
1762 848 Pnt
1079 414 Pnt
551 953 Pnt
1191 393 Pnt
1808 907 Pnt
1667 741 Pnt
798 666 Pnt
866 596 Pnt
497 1044 Pnt
1104 406 Pnt
625 851 Pnt
1863 993 Pnt
812 651 Pnt
1636 707 Pnt
1470 533 Pnt
1840 957 Pnt
1382 451 Pnt
795 669 Pnt
1558 623 Pnt
1432 495 Pnt
902 559 Pnt
1528 591 Pnt
1509 571 Pnt
498 1042 Pnt
400 1276 Pnt
1908 1076 Pnt
472 1091 Pnt
1902 1065 Pnt
1587 654 Pnt
432 1182 Pnt
824 638 Pnt
659 815 Pnt
1334 418 Pnt
1261 398 Pnt
1206 393 Pnt
1408 474 Pnt
1860 988 Pnt
646 829 Pnt
1751 837 Pnt
831 632 Pnt
1954 1181 Pnt
1475 537 Pnt
814 649 Pnt
1420 485 Pnt
686 785 Pnt
1639 710 Pnt
1061 420 Pnt
1244 396 Pnt
1121 402 Pnt
441 1159 Pnt
451 1138 Pnt
1131 400 Pnt
476 1083 Pnt
1965 1210 Pnt
1017 453 Pnt
1084 412 Pnt
1667 742 Pnt
785 680 Pnt
425 1200 Pnt
1122 402 Pnt
602 879 Pnt
1071 417 Pnt
1545 609 Pnt
1508 570 Pnt
1527 590 Pnt
1141 398 Pnt
1805 902 Pnt
1295 406 Pnt
1087 411 Pnt
1204 393 Pnt
1741 825 Pnt
1054 424 Pnt
1070 417 Pnt
771 694 Pnt
814 649 Pnt
664 809 Pnt
729 739 Pnt
1831 942 Pnt
1147 397 Pnt
1285 403 Pnt
702 768 Pnt
1793 886 Pnt
1701 779 Pnt
1184 393 Pnt
1013 456 Pnt
630 845 Pnt
1136 399 Pnt
737 731 Pnt
639 836 Pnt
674 798 Pnt
459 1118 Pnt
569 926 Pnt
1540 604 Pnt
1496 558 Pnt
1548 612 Pnt
1915 1091 Pnt
1394 461 Pnt
827 635 Pnt
947 516 Pnt
446 1147 Pnt
727 741 Pnt
1991 1297 Pnt
1989 1289 Pnt
534 980 Pnt
1789 880 Pnt
1628 698 Pnt
701 770 Pnt
1147 397 Pnt
1985 1273 Pnt
1126 401 Pnt
413 1236 Pnt
709 761 Pnt
962 502 Pnt
511 1018 Pnt
971 493 Pnt
541 969 Pnt
1667 742 Pnt
1899 1059 Pnt
603 877 Pnt
1056 423 Pnt
777 688 Pnt
858 604 Pnt
1768 854 Pnt
1595 662 Pnt
855 606 Pnt
1434 498 Pnt
862 599 Pnt
1094 409 Pnt
1073 416 Pnt
1598 665 Pnt
1829 938 Pnt
713 757 Pnt
1446 509 Pnt
1001 466 Pnt
591 894 Pnt
1758 844 Pnt
1866 999 Pnt
1245 396 Pnt
1749 834 Pnt
1976 1245 Pnt
1079 414 Pnt
1914 1088 Pnt
1995 1310 Pnt
1624 694 Pnt
1005 463 Pnt
1659 732 Pnt
1706 785 Pnt
1387 456 Pnt
1042 433 Pnt
557 943 Pnt
1619 689 Pnt
743 724 Pnt
1018 452 Pnt
1244 396 Pnt
1550 614 Pnt
1199 393 Pnt
953 510 Pnt
704 766 Pnt
782 683 Pnt
1928 1118 Pnt
1825 932 Pnt
1827 935 Pnt
1029 443 Pnt
777 687 Pnt
1340 420 Pnt
523 998 Pnt
1915 1090 Pnt
687 784 Pnt
465 1105 Pnt
531 985 Pnt
809 654 Pnt
1696 774 Pnt
840 622 Pnt
601 879 Pnt
1722 804 Pnt
1747 832 Pnt
495 1047 Pnt
544 965 Pnt
1968 1220 Pnt
1520 583 Pnt
1162 395 Pnt
982 483 Pnt
1347 425 Pnt
1710 789 Pnt
808 655 Pnt
1728 810 Pnt
1427 491 Pnt
687 785 Pnt
1688 765 Pnt
1166 395 Pnt
1473 536 Pnt
1050 427 Pnt
1449 512 Pnt
1175 394 Pnt
1708 787 Pnt
1769 856 Pnt
1982 1264 Pnt
492 1052 Pnt
1662 736 Pnt
485 1065 Pnt
1114 404 Pnt
1736 820 Pnt
1359 433 Pnt
1457 520 Pnt
705 765 Pnt
1879 1020 Pnt
1803 901 Pnt
676 797 Pnt
581 908 Pnt
855 607 Pnt
1948 1167 Pnt
1016 454 Pnt
1121 402 Pnt
874 587 Pnt
1801 897 Pnt
1628 699 Pnt
1206 393 Pnt
1318 413 Pnt
1.000 UL
LT1
400 1089 M
16 0 V
17 0 V
16 0 V
16 0 V
16 0 V
17 0 V
16 0 V
16 0 V
16 0 V
17 0 V
16 0 V
16 0 V
16 0 V
17 0 V
16 0 V
16 0 V
16 0 V
17 0 V
16 0 V
16 0 V
17 0 V
16 0 V
16 0 V
16 0 V
17 0 V
16 0 V
16 0 V
16 0 V
17 0 V
16 0 V
16 0 V
16 0 V
17 0 V
16 0 V
16 0 V
16 0 V
17 0 V
16 0 V
16 0 V
17 0 V
16 0 V
16 0 V
16 0 V
17 0 V
16 0 V
16 0 V
16 0 V
17 0 V
16 0 V
16 0 V
16 0 V
17 0 V
16 0 V
16 0 V
16 0 V
17 0 V
16 0 V
16 0 V
16 0 V
17 0 V
16 0 V
16 0 V
17 0 V
16 0 V
16 0 V
16 0 V
17 0 V
16 0 V
16 0 V
16 0 V
17 0 V
16 0 V
16 0 V
16 0 V
17 0 V
16 0 V
16 0 V
16 0 V
17 0 V
16 0 V
16 0 V
17 0 V
16 0 V
16 0 V
16 0 V
17 0 V
16 0 V
16 0 V
16 0 V
17 0 V
16 0 V
16 0 V
16 0 V
17 0 V
16 0 V
16 0 V
16 0 V
17 0 V
16 0 V
1.000 UL
LT2
400 712 M
16 0 V
17 0 V
16 0 V
16 0 V
16 0 V
17 0 V
16 0 V
16 0 V
16 0 V
17 0 V
16 0 V
16 0 V
16 0 V
17 0 V
16 0 V
16 0 V
16 0 V
17 0 V
16 0 V
16 0 V
17 0 V
16 0 V
16 0 V
16 0 V
17 0 V
16 0 V
16 0 V
16 0 V
17 0 V
16 0 V
16 0 V
16 0 V
17 0 V
16 0 V
16 0 V
16 0 V
17 0 V
16 0 V
16 0 V
17 0 V
16 0 V
16 0 V
16 0 V
17 0 V
16 0 V
16 0 V
16 0 V
17 0 V
16 0 V
16 0 V
16 0 V
17 0 V
16 0 V
16 0 V
16 0 V
17 0 V
16 0 V
16 0 V
16 0 V
17 0 V
16 0 V
16 0 V
17 0 V
16 0 V
16 0 V
16 0 V
17 0 V
16 0 V
16 0 V
16 0 V
17 0 V
16 0 V
16 0 V
16 0 V
17 0 V
16 0 V
16 0 V
16 0 V
17 0 V
16 0 V
16 0 V
17 0 V
16 0 V
16 0 V
16 0 V
17 0 V
16 0 V
16 0 V
16 0 V
17 0 V
16 0 V
16 0 V
16 0 V
17 0 V
16 0 V
16 0 V
16 0 V
17 0 V
16 0 V
0.500 UL
LTb
400 300 M
1610 0 V
0 1026 V
-1610 0 V
400 300 L
1.000 UP
stroke
grestore
end
showpage
}}%
\put(212,712){\makebox(0,0)[l]{$1/6$}}%
\put(212,1089){\makebox(0,0)[l]{$1/2$}}%
\put(1205,50){\makebox(0,0){$\mu$ [TeV]}}%
\put(100,813){%
\special{ps: gsave currentpoint currentpoint translate
270 rotate neg exch neg exch translate}%
\makebox(0,0)[b]{\shortstack{$|x|$}}%
\special{ps: currentpoint grestore moveto}%
}%
\put(2010,200){\makebox(0,0){ 1.5}}%
\put(1742,200){\makebox(0,0){ 1}}%
\put(1473,200){\makebox(0,0){ 0.5}}%
\put(1205,200){\makebox(0,0){ 0}}%
\put(937,200){\makebox(0,0){-0.5}}%
\put(668,200){\makebox(0,0){-1}}%
\put(400,200){\makebox(0,0){-1.5}}%
\put(350,1326){\makebox(0,0)[r]{ 1}}%
\put(350,537){\makebox(0,0)[r]{ 0.1}}%
\end{picture}%
\endgroup
 

%% file: schuessler_susy07fig3.tex
\begingroup%
  \makeatletter%
  \newcommand{\GNUPLOTspecial}{%
    \@sanitize\catcode`\%=14\relax\special}%
  \setlength{\unitlength}{0.1bp}%
\begin{picture}(2160,1438)(0,0)%
{\GNUPLOTspecial{"
/gnudict 256 dict def
gnudict begin
/Color false def
/Solid false def
/gnulinewidth 5.000 def
/userlinewidth gnulinewidth def
/vshift -33 def
/dl {10.0 mul} def
/hpt_ 31.5 def
/vpt_ 31.5 def
/hpt hpt_ def
/vpt vpt_ def
/Rounded false def
/M {moveto} bind def
/L {lineto} bind def
/R {rmoveto} bind def
/V {rlineto} bind def
/N {newpath moveto} bind def
/C {setrgbcolor} bind def
/f {rlineto fill} bind def
/vpt2 vpt 2 mul def
/hpt2 hpt 2 mul def
/Lshow { currentpoint stroke M
  0 vshift R show } def
/Rshow { currentpoint stroke M
  dup stringwidth pop neg vshift R show } def
/Cshow { currentpoint stroke M
  dup stringwidth pop -2 div vshift R show } def
/UP { dup vpt_ mul /vpt exch def hpt_ mul /hpt exch def
  /hpt2 hpt 2 mul def /vpt2 vpt 2 mul def } def
/DL { Color {setrgbcolor Solid {pop []} if 0 setdash }
 {pop pop pop 0 setgray Solid {pop []} if 0 setdash} ifelse } def
/BL { stroke userlinewidth 2 mul setlinewidth
      Rounded { 1 setlinejoin 1 setlinecap } if } def
/AL { stroke userlinewidth 2 div setlinewidth
      Rounded { 1 setlinejoin 1 setlinecap } if } def
/UL { dup gnulinewidth mul /userlinewidth exch def
      dup 1 lt {pop 1} if 10 mul /udl exch def } def
/PL { stroke userlinewidth setlinewidth
      Rounded { 1 setlinejoin 1 setlinecap } if } def
/LTw { PL [] 1 setgray } def
/LTb { BL [] 0 0 0 DL } def
/LTa { AL [1 udl mul 2 udl mul] 0 setdash 0 0 0 setrgbcolor } def
/LT0 { PL [] 1 0 0 DL } def
/LT1 { PL [4 dl 2 dl] 0 1 0 DL } def
/LT2 { PL [2 dl 3 dl] 0 0 1 DL } def
/LT3 { PL [1 dl 1.5 dl] 1 0 1 DL } def
/LT4 { PL [5 dl 2 dl 1 dl 2 dl] 0 1 1 DL } def
/LT5 { PL [4 dl 3 dl 1 dl 3 dl] 1 1 0 DL } def
/LT6 { PL [2 dl 2 dl 2 dl 4 dl] 0 0 0 DL } def
/LT7 { PL [2 dl 2 dl 2 dl 2 dl 2 dl 4 dl] 1 0.3 0 DL } def
/LT8 { PL [2 dl 2 dl 2 dl 2 dl 2 dl 2 dl 2 dl 4 dl] 0.5 0.5 0.5 DL } def
/Pnt { stroke [] 0 setdash
   gsave 1 setlinecap M 0 0 V stroke grestore } def
/Dia { stroke [] 0 setdash 2 copy vpt add M
  hpt neg vpt neg V hpt vpt neg V
  hpt vpt V hpt neg vpt V closepath stroke
  Pnt } def
/Pls { stroke [] 0 setdash vpt sub M 0 vpt2 V
  currentpoint stroke M
  hpt neg vpt neg R hpt2 0 V stroke
  } def
/Box { stroke [] 0 setdash 2 copy exch hpt sub exch vpt add M
  0 vpt2 neg V hpt2 0 V 0 vpt2 V
  hpt2 neg 0 V closepath stroke
  Pnt } def
/Crs { stroke [] 0 setdash exch hpt sub exch vpt add M
  hpt2 vpt2 neg V currentpoint stroke M
  hpt2 neg 0 R hpt2 vpt2 V stroke } def
/TriU { stroke [] 0 setdash 2 copy vpt 1.12 mul add M
  hpt neg vpt -1.62 mul V
  hpt 2 mul 0 V
  hpt neg vpt 1.62 mul V closepath stroke
  Pnt  } def
/Star { 2 copy Pls Crs } def
/BoxF { stroke [] 0 setdash exch hpt sub exch vpt add M
  0 vpt2 neg V  hpt2 0 V  0 vpt2 V
  hpt2 neg 0 V  closepath fill } def
/TriUF { stroke [] 0 setdash vpt 1.12 mul add M
  hpt neg vpt -1.62 mul V
  hpt 2 mul 0 V
  hpt neg vpt 1.62 mul V closepath fill } def
/TriD { stroke [] 0 setdash 2 copy vpt 1.12 mul sub M
  hpt neg vpt 1.62 mul V
  hpt 2 mul 0 V
  hpt neg vpt -1.62 mul V closepath stroke
  Pnt  } def
/TriDF { stroke [] 0 setdash vpt 1.12 mul sub M
  hpt neg vpt 1.62 mul V
  hpt 2 mul 0 V
  hpt neg vpt -1.62 mul V closepath fill} def
/DiaF { stroke [] 0 setdash vpt add M
  hpt neg vpt neg V hpt vpt neg V
  hpt vpt V hpt neg vpt V closepath fill } def
/Pent { stroke [] 0 setdash 2 copy gsave
  translate 0 hpt M 4 {72 rotate 0 hpt L} repeat
  closepath stroke grestore Pnt } def
/PentF { stroke [] 0 setdash gsave
  translate 0 hpt M 4 {72 rotate 0 hpt L} repeat
  closepath fill grestore } def
/Circle { stroke [] 0 setdash 2 copy
  hpt 0 360 arc stroke Pnt } def
/CircleF { stroke [] 0 setdash hpt 0 360 arc fill } def
/C0 { BL [] 0 setdash 2 copy moveto vpt 90 450  arc } bind def
/C1 { BL [] 0 setdash 2 copy        moveto
       2 copy  vpt 0 90 arc closepath fill
               vpt 0 360 arc closepath } bind def
/C2 { BL [] 0 setdash 2 copy moveto
       2 copy  vpt 90 180 arc closepath fill
               vpt 0 360 arc closepath } bind def
/C3 { BL [] 0 setdash 2 copy moveto
       2 copy  vpt 0 180 arc closepath fill
               vpt 0 360 arc closepath } bind def
/C4 { BL [] 0 setdash 2 copy moveto
       2 copy  vpt 180 270 arc closepath fill
               vpt 0 360 arc closepath } bind def
/C5 { BL [] 0 setdash 2 copy moveto
       2 copy  vpt 0 90 arc
       2 copy moveto
       2 copy  vpt 180 270 arc closepath fill
               vpt 0 360 arc } bind def
/C6 { BL [] 0 setdash 2 copy moveto
      2 copy  vpt 90 270 arc closepath fill
              vpt 0 360 arc closepath } bind def
/C7 { BL [] 0 setdash 2 copy moveto
      2 copy  vpt 0 270 arc closepath fill
              vpt 0 360 arc closepath } bind def
/C8 { BL [] 0 setdash 2 copy moveto
      2 copy vpt 270 360 arc closepath fill
              vpt 0 360 arc closepath } bind def
/C9 { BL [] 0 setdash 2 copy moveto
      2 copy  vpt 270 450 arc closepath fill
              vpt 0 360 arc closepath } bind def
/C10 { BL [] 0 setdash 2 copy 2 copy moveto vpt 270 360 arc closepath fill
       2 copy moveto
       2 copy vpt 90 180 arc closepath fill
               vpt 0 360 arc closepath } bind def
/C11 { BL [] 0 setdash 2 copy moveto
       2 copy  vpt 0 180 arc closepath fill
       2 copy moveto
       2 copy  vpt 270 360 arc closepath fill
               vpt 0 360 arc closepath } bind def
/C12 { BL [] 0 setdash 2 copy moveto
       2 copy  vpt 180 360 arc closepath fill
               vpt 0 360 arc closepath } bind def
/C13 { BL [] 0 setdash  2 copy moveto
       2 copy  vpt 0 90 arc closepath fill
       2 copy moveto
       2 copy  vpt 180 360 arc closepath fill
               vpt 0 360 arc closepath } bind def
/C14 { BL [] 0 setdash 2 copy moveto
       2 copy  vpt 90 360 arc closepath fill
               vpt 0 360 arc } bind def
/C15 { BL [] 0 setdash 2 copy vpt 0 360 arc closepath fill
               vpt 0 360 arc closepath } bind def
/Rec   { newpath 4 2 roll moveto 1 index 0 rlineto 0 exch rlineto
       neg 0 rlineto closepath } bind def
/Square { dup Rec } bind def
/Bsquare { vpt sub exch vpt sub exch vpt2 Square } bind def
/S0 { BL [] 0 setdash 2 copy moveto 0 vpt rlineto BL Bsquare } bind def
/S1 { BL [] 0 setdash 2 copy vpt Square fill Bsquare } bind def
/S2 { BL [] 0 setdash 2 copy exch vpt sub exch vpt Square fill Bsquare } bind def
/S3 { BL [] 0 setdash 2 copy exch vpt sub exch vpt2 vpt Rec fill Bsquare } bind def
/S4 { BL [] 0 setdash 2 copy exch vpt sub exch vpt sub vpt Square fill Bsquare } bind def
/S5 { BL [] 0 setdash 2 copy 2 copy vpt Square fill
       exch vpt sub exch vpt sub vpt Square fill Bsquare } bind def
/S6 { BL [] 0 setdash 2 copy exch vpt sub exch vpt sub vpt vpt2 Rec fill Bsquare } bind def
/S7 { BL [] 0 setdash 2 copy exch vpt sub exch vpt sub vpt vpt2 Rec fill
       2 copy vpt Square fill
       Bsquare } bind def
/S8 { BL [] 0 setdash 2 copy vpt sub vpt Square fill Bsquare } bind def
/S9 { BL [] 0 setdash 2 copy vpt sub vpt vpt2 Rec fill Bsquare } bind def
/S10 { BL [] 0 setdash 2 copy vpt sub vpt Square fill 2 copy exch vpt sub exch vpt Square fill
       Bsquare } bind def
/S11 { BL [] 0 setdash 2 copy vpt sub vpt Square fill 2 copy exch vpt sub exch vpt2 vpt Rec fill
       Bsquare } bind def
/S12 { BL [] 0 setdash 2 copy exch vpt sub exch vpt sub vpt2 vpt Rec fill Bsquare } bind def
/S13 { BL [] 0 setdash 2 copy exch vpt sub exch vpt sub vpt2 vpt Rec fill
       2 copy vpt Square fill Bsquare } bind def
/S14 { BL [] 0 setdash 2 copy exch vpt sub exch vpt sub vpt2 vpt Rec fill
       2 copy exch vpt sub exch vpt Square fill Bsquare } bind def
/S15 { BL [] 0 setdash 2 copy Bsquare fill Bsquare } bind def
/D0 { gsave translate 45 rotate 0 0 S0 stroke grestore } bind def
/D1 { gsave translate 45 rotate 0 0 S1 stroke grestore } bind def
/D2 { gsave translate 45 rotate 0 0 S2 stroke grestore } bind def
/D3 { gsave translate 45 rotate 0 0 S3 stroke grestore } bind def
/D4 { gsave translate 45 rotate 0 0 S4 stroke grestore } bind def
/D5 { gsave translate 45 rotate 0 0 S5 stroke grestore } bind def
/D6 { gsave translate 45 rotate 0 0 S6 stroke grestore } bind def
/D7 { gsave translate 45 rotate 0 0 S7 stroke grestore } bind def
/D8 { gsave translate 45 rotate 0 0 S8 stroke grestore } bind def
/D9 { gsave translate 45 rotate 0 0 S9 stroke grestore } bind def
/D10 { gsave translate 45 rotate 0 0 S10 stroke grestore } bind def
/D11 { gsave translate 45 rotate 0 0 S11 stroke grestore } bind def
/D12 { gsave translate 45 rotate 0 0 S12 stroke grestore } bind def
/D13 { gsave translate 45 rotate 0 0 S13 stroke grestore } bind def
/D14 { gsave translate 45 rotate 0 0 S14 stroke grestore } bind def
/D15 { gsave translate 45 rotate 0 0 S15 stroke grestore } bind def
/DiaE { stroke [] 0 setdash vpt add M
  hpt neg vpt neg V hpt vpt neg V
  hpt vpt V hpt neg vpt V closepath stroke } def
/BoxE { stroke [] 0 setdash exch hpt sub exch vpt add M
  0 vpt2 neg V hpt2 0 V 0 vpt2 V
  hpt2 neg 0 V closepath stroke } def
/TriUE { stroke [] 0 setdash vpt 1.12 mul add M
  hpt neg vpt -1.62 mul V
  hpt 2 mul 0 V
  hpt neg vpt 1.62 mul V closepath stroke } def
/TriDE { stroke [] 0 setdash vpt 1.12 mul sub M
  hpt neg vpt 1.62 mul V
  hpt 2 mul 0 V
  hpt neg vpt -1.62 mul V closepath stroke } def
/PentE { stroke [] 0 setdash gsave
  translate 0 hpt M 4 {72 rotate 0 hpt L} repeat
  closepath stroke grestore } def
/CircE { stroke [] 0 setdash 
  hpt 0 360 arc stroke } def
/Opaque { gsave closepath 1 setgray fill grestore 0 setgray closepath } def
/DiaW { stroke [] 0 setdash vpt add M
  hpt neg vpt neg V hpt vpt neg V
  hpt vpt V hpt neg vpt V Opaque stroke } def
/BoxW { stroke [] 0 setdash exch hpt sub exch vpt add M
  0 vpt2 neg V hpt2 0 V 0 vpt2 V
  hpt2 neg 0 V Opaque stroke } def
/TriUW { stroke [] 0 setdash vpt 1.12 mul add M
  hpt neg vpt -1.62 mul V
  hpt 2 mul 0 V
  hpt neg vpt 1.62 mul V Opaque stroke } def
/TriDW { stroke [] 0 setdash vpt 1.12 mul sub M
  hpt neg vpt 1.62 mul V
  hpt 2 mul 0 V
  hpt neg vpt -1.62 mul V Opaque stroke } def
/PentW { stroke [] 0 setdash gsave
  translate 0 hpt M 4 {72 rotate 0 hpt L} repeat
  Opaque stroke grestore } def
/CircW { stroke [] 0 setdash 
  hpt 0 360 arc Opaque stroke } def
/BoxFill { gsave Rec 1 setgray fill grestore } def
/BoxColFill {
  gsave Rec
  /Fillden exch def
  currentrgbcolor
  /ColB exch def /ColG exch def /ColR exch def
  /ColR ColR Fillden mul Fillden sub 1 add def
  /ColG ColG Fillden mul Fillden sub 1 add def
  /ColB ColB Fillden mul Fillden sub 1 add def
  ColR ColG ColB setrgbcolor
  fill grestore } def
%
%
/PatternFill { gsave /PFa [ 9 2 roll ] def
    PFa 0 get PFa 2 get 2 div add PFa 1 get PFa 3 get 2 div add translate
    PFa 2 get -2 div PFa 3 get -2 div PFa 2 get PFa 3 get Rec
    gsave 1 setgray fill grestore clip
    currentlinewidth 0.5 mul setlinewidth
    /PFs PFa 2 get dup mul PFa 3 get dup mul add sqrt def
    0 0 M PFa 5 get rotate PFs -2 div dup translate
	0 1 PFs PFa 4 get div 1 add floor cvi
	{ PFa 4 get mul 0 M 0 PFs V } for
    0 PFa 6 get ne {
	0 1 PFs PFa 4 get div 1 add floor cvi
	{ PFa 4 get mul 0 2 1 roll M PFs 0 V } for
    } if
    stroke grestore } def
/Symbol-Oblique /Symbol findfont [1 0 .167 1 0 0] makefont
dup length dict begin {1 index /FID eq {pop pop} {def} ifelse} forall
currentdict end definefont pop
end
gnudict begin
gsave
0 0 translate
0.100 0.100 scale
0 setgray
newpath
0.500 UL
LTb
400 300 M
37 0 V
1573 0 R
-37 0 V
0.500 UL
LTb
400 300 M
18 0 V
1592 0 R
-18 0 V
400 819 M
37 0 V
1573 0 R
-37 0 V
0.500 UL
LTb
400 300 M
18 0 V
1592 0 R
-18 0 V
400 456 M
18 0 V
1592 0 R
-18 0 V
400 548 M
18 0 V
1592 0 R
-18 0 V
400 613 M
18 0 V
1592 0 R
-18 0 V
400 663 M
18 0 V
1592 0 R
-18 0 V
400 704 M
18 0 V
1592 0 R
-18 0 V
400 739 M
18 0 V
1592 0 R
-18 0 V
400 769 M
18 0 V
1592 0 R
-18 0 V
400 796 M
18 0 V
1592 0 R
-18 0 V
400 819 M
18 0 V
1592 0 R
-18 0 V
400 1339 M
37 0 V
1573 0 R
-37 0 V
0.500 UL
LTb
400 819 M
18 0 V
1592 0 R
-18 0 V
400 976 M
18 0 V
1592 0 R
-18 0 V
400 1067 M
18 0 V
1592 0 R
-18 0 V
400 1132 M
18 0 V
1592 0 R
-18 0 V
400 1183 M
18 0 V
1592 0 R
-18 0 V
400 1224 M
18 0 V
1592 0 R
-18 0 V
400 1259 M
18 0 V
1592 0 R
-18 0 V
400 1289 M
18 0 V
1592 0 R
-18 0 V
400 1315 M
18 0 V
1592 0 R
-18 0 V
400 1339 M
18 0 V
1592 0 R
-18 0 V
400 300 M
0 18 V
0 1021 R
0 -18 V
458 300 M
0 18 V
0 1021 R
0 -18 V
515 300 M
0 37 V
0 1002 R
0 -37 V
0.500 UL
LTb
573 300 M
0 18 V
0 1021 R
0 -18 V
630 300 M
0 18 V
0 1021 R
0 -18 V
688 300 M
0 18 V
0 1021 R
0 -18 V
745 300 M
0 37 V
0 1002 R
0 -37 V
0.500 UL
LTb
803 300 M
0 18 V
0 1021 R
0 -18 V
860 300 M
0 18 V
0 1021 R
0 -18 V
918 300 M
0 18 V
0 1021 R
0 -18 V
975 300 M
0 37 V
0 1002 R
0 -37 V
0.500 UL
LTb
1033 300 M
0 18 V
0 1021 R
0 -18 V
1090 300 M
0 18 V
0 1021 R
0 -18 V
1148 300 M
0 18 V
0 1021 R
0 -18 V
1205 300 M
0 37 V
0 1002 R
0 -37 V
0.500 UL
LTb
1263 300 M
0 18 V
0 1021 R
0 -18 V
1320 300 M
0 18 V
0 1021 R
0 -18 V
1378 300 M
0 18 V
0 1021 R
0 -18 V
1435 300 M
0 37 V
0 1002 R
0 -37 V
0.500 UL
LTb
1493 300 M
0 18 V
0 1021 R
0 -18 V
1550 300 M
0 18 V
0 1021 R
0 -18 V
1608 300 M
0 18 V
0 1021 R
0 -18 V
1665 300 M
0 37 V
0 1002 R
0 -37 V
0.500 UL
LTb
1723 300 M
0 18 V
0 1021 R
0 -18 V
1780 300 M
0 18 V
0 1021 R
0 -18 V
1838 300 M
0 18 V
0 1021 R
0 -18 V
1895 300 M
0 37 V
0 1002 R
0 -37 V
0.500 UL
LTb
1953 300 M
0 18 V
0 1021 R
0 -18 V
2010 300 M
0 18 V
0 1021 R
0 -18 V
0.500 UL
LTb
400 300 M
1610 0 V
0 1039 V
-1610 0 V
400 300 L
LTb
LTb
1.000 UP
LTb
LTb
1.000 UL
LT2
628 300 M
0 1039 V
1.000 UL
LT2
1782 300 M
0 1039 V
1.000 UL
LT0
1823 1162 Pnt
1730 975 Pnt
1178 686 Pnt
1270 670 Pnt
1286 660 Pnt
709 905 Pnt
734 845 Pnt
521 1272 Pnt
1447 479 Pnt
886 574 Pnt
1618 720 Pnt
540 1241 Pnt
687 956 Pnt
1072 614 Pnt
1882 1263 Pnt
807 694 Pnt
1562 633 Pnt
1198 688 Pnt
1210 688 Pnt
539 1243 Pnt
713 896 Pnt
815 680 Pnt
1415 498 Pnt
1900 1291 Pnt
993 497 Pnt
1396 526 Pnt
1458 490 Pnt
961 479 Pnt
1740 996 Pnt
1050 585 Pnt
1150 676 Pnt
787 727 Pnt
937 505 Pnt
685 961 Pnt
1142 672 Pnt
740 833 Pnt
1476 511 Pnt
1501 544 Pnt
599 1137 Pnt
1684 868 Pnt
1413 502 Pnt
815 680 Pnt
1897 1287 Pnt
1155 678 Pnt
1534 592 Pnt
947 494 Pnt
1757 1033 Pnt
1148 675 Pnt
1237 684 Pnt
1838 1190 Pnt
729 858 Pnt
1561 630 Pnt
1857 1223 Pnt
1931 1338 Pnt
1650 789 Pnt
1147 675 Pnt
1560 629 Pnt
751 807 Pnt
1268 672 Pnt
1230 686 Pnt
1537 596 Pnt
907 545 Pnt
696 935 Pnt
1886 1269 Pnt
1372 565 Pnt
549 1225 Pnt
1129 665 Pnt
1326 624 Pnt
1834 1182 Pnt
658 1019 Pnt
1906 1301 Pnt
919 529 Pnt
1562 633 Pnt
1067 607 Pnt
1753 1025 Pnt
1490 529 Pnt
1587 671 Pnt
1205 688 Pnt
1056 593 Pnt
1355 589 Pnt
1299 651 Pnt
1101 643 Pnt
1657 805 Pnt
1927 1332 Pnt
841 640 Pnt
683 964 Pnt
780 741 Pnt
982 485 Pnt
601 1134 Pnt
1635 754 Pnt
1382 549 Pnt
1899 1290 Pnt
1441 477 Pnt
1157 679 Pnt
975 479 Pnt
853 622 Pnt
1589 673 Pnt
571 1188 Pnt
1309 642 Pnt
1490 530 Pnt
1339 610 Pnt
1418 494 Pnt
938 505 Pnt
881 582 Pnt
652 1032 Pnt
895 562 Pnt
703 919 Pnt
1072 613 Pnt
1415 499 Pnt
942 499 Pnt
1105 647 Pnt
517 1277 Pnt
1743 1003 Pnt
981 484 Pnt
1246 681 Pnt
1669 833 Pnt
693 941 Pnt
1373 563 Pnt
947 494 Pnt
1186 687 Pnt
1488 527 Pnt
1418 495 Pnt
552 1220 Pnt
1409 507 Pnt
982 485 Pnt
592 1151 Pnt
1287 659 Pnt
840 642 Pnt
631 1076 Pnt
1281 663 Pnt
539 1242 Pnt
1210 688 Pnt
834 651 Pnt
1334 616 Pnt
1343 605 Pnt
1115 655 Pnt
1306 644 Pnt
688 954 Pnt
1249 680 Pnt
896 560 Pnt
1067 607 Pnt
611 1115 Pnt
1692 888 Pnt
1684 869 Pnt
1430 482 Pnt
1774 1069 Pnt
1889 1275 Pnt
1598 688 Pnt
789 723 Pnt
1749 1017 Pnt
1772 1064 Pnt
648 1040 Pnt
959 481 Pnt
1166 682 Pnt
1306 644 Pnt
558 1210 Pnt
1579 658 Pnt
975 479 Pnt
1795 1109 Pnt
650 1036 Pnt
1862 1230 Pnt
1735 985 Pnt
1628 739 Pnt
1348 599 Pnt
941 501 Pnt
1215 688 Pnt
799 706 Pnt
1319 632 Pnt
522 1271 Pnt
553 1220 Pnt
814 682 Pnt
725 866 Pnt
1138 670 Pnt
1918 1319 Pnt
514 1283 Pnt
921 525 Pnt
1827 1170 Pnt
1186 687 Pnt
964 478 Pnt
1396 527 Pnt
1516 565 Pnt
1287 660 Pnt
513 1284 Pnt
1413 501 Pnt
1396 528 Pnt
1435 479 Pnt
1113 653 Pnt
1511 558 Pnt
1551 616 Pnt
1423 489 Pnt
1346 601 Pnt
1827 1170 Pnt
848 630 Pnt
829 659 Pnt
688 953 Pnt
1216 688 Pnt
1404 515 Pnt
869 599 Pnt
1575 652 Pnt
1005 515 Pnt
911 540 Pnt
499 1305 Pnt
1551 616 Pnt
1898 1287 Pnt
943 498 Pnt
582 1168 Pnt
1755 1029 Pnt
1692 889 Pnt
592 1151 Pnt
1664 822 Pnt
1874 1250 Pnt
621 1096 Pnt
961 480 Pnt
786 729 Pnt
1014 530 Pnt
1584 666 Pnt
1014 530 Pnt
1125 662 Pnt
1367 572 Pnt
1911 1308 Pnt
1920 1322 Pnt
677 977 Pnt
767 768 Pnt
481 1332 Pnt
1769 1058 Pnt
1377 557 Pnt
1162 681 Pnt
1514 563 Pnt
1392 534 Pnt
1838 1190 Pnt
1717 945 Pnt
791 719 Pnt
1564 636 Pnt
1440 477 Pnt
1923 1326 Pnt
956 485 Pnt
1874 1251 Pnt
1581 661 Pnt
1714 939 Pnt
541 1239 Pnt
801 702 Pnt
1428 484 Pnt
733 849 Pnt
1508 554 Pnt
1483 520 Pnt
1825 1166 Pnt
910 541 Pnt
1449 481 Pnt
527 1262 Pnt
1137 669 Pnt
1475 510 Pnt
568 1193 Pnt
1761 1041 Pnt
1420 492 Pnt
1771 1063 Pnt
697 934 Pnt
1374 561 Pnt
564 1200 Pnt
1880 1260 Pnt
1908 1304 Pnt
983 486 Pnt
1504 548 Pnt
1284 662 Pnt
730 856 Pnt
1089 631 Pnt
680 973 Pnt
1216 688 Pnt
655 1025 Pnt
866 604 Pnt
1452 484 Pnt
1193 688 Pnt
941 501 Pnt
845 635 Pnt
694 940 Pnt
984 487 Pnt
545 1232 Pnt
974 479 Pnt
1133 667 Pnt
511 1287 Pnt
1528 582 Pnt
726 866 Pnt
935 508 Pnt
1237 684 Pnt
824 666 Pnt
1719 951 Pnt
1629 742 Pnt
1068 609 Pnt
1230 686 Pnt
1700 908 Pnt
1063 602 Pnt
1306 644 Pnt
1447 479 Pnt
1305 646 Pnt
1397 526 Pnt
1585 667 Pnt
1748 1013 Pnt
926 519 Pnt
1238 684 Pnt
1311 640 Pnt
1885 1267 Pnt
511 1287 Pnt
1737 990 Pnt
1082 625 Pnt
1071 612 Pnt
1618 721 Pnt
1608 703 Pnt
822 669 Pnt
1192 688 Pnt
1847 1204 Pnt
536 1247 Pnt
1557 625 Pnt
1388 539 Pnt
852 624 Pnt
500 1304 Pnt
1188 687 Pnt
1754 1028 Pnt
1899 1290 Pnt
1192 688 Pnt
869 600 Pnt
1800 1119 Pnt
1678 855 Pnt
655 1027 Pnt
1020 540 Pnt
589 1156 Pnt
1605 699 Pnt
709 906 Pnt
888 571 Pnt
988 492 Pnt
1287 660 Pnt
773 755 Pnt
1500 542 Pnt
1519 569 Pnt
997 504 Pnt
1605 698 Pnt
660 1014 Pnt
532 1255 Pnt
564 1201 Pnt
514 1282 Pnt
888 572 Pnt
1219 687 Pnt
953 487 Pnt
1795 1110 Pnt
1373 563 Pnt
1565 637 Pnt
487 1323 Pnt
1528 583 Pnt
703 920 Pnt
1817 1151 Pnt
997 503 Pnt
1520 571 Pnt
1071 612 Pnt
535 1250 Pnt
776 748 Pnt
754 798 Pnt
1863 1232 Pnt
1088 630 Pnt
1402 518 Pnt
847 632 Pnt
1406 511 Pnt
505 1297 Pnt
793 715 Pnt
1166 682 Pnt
1212 688 Pnt
1158 680 Pnt
1178 685 Pnt
1663 819 Pnt
1004 513 Pnt
1468 502 Pnt
526 1263 Pnt
1056 593 Pnt
1047 580 Pnt
1295 654 Pnt
500 1304 Pnt
1082 625 Pnt
1418 495 Pnt
1925 1330 Pnt
1851 1211 Pnt
1017 534 Pnt
1331 619 Pnt
727 863 Pnt
608 1120 Pnt
1379 554 Pnt
690 950 Pnt
1211 688 Pnt
720 878 Pnt
554 1218 Pnt
667 999 Pnt
971 478 Pnt
1461 493 Pnt
1071 612 Pnt
1248 680 Pnt
1386 543 Pnt
1420 492 Pnt
685 960 Pnt
1358 585 Pnt
1339 610 Pnt
484 1329 Pnt
1042 573 Pnt
545 1232 Pnt
951 490 Pnt
1193 688 Pnt
1396 527 Pnt
1118 657 Pnt
604 1128 Pnt
971 478 Pnt
798 707 Pnt
1000 508 Pnt
530 1257 Pnt
893 564 Pnt
1555 622 Pnt
1365 575 Pnt
1918 1318 Pnt
614 1110 Pnt
1767 1054 Pnt
1383 548 Pnt
1510 557 Pnt
1774 1068 Pnt
886 575 Pnt
1112 653 Pnt
501 1303 Pnt
1304 646 Pnt
1846 1203 Pnt
1150 676 Pnt
1279 665 Pnt
585 1163 Pnt
577 1178 Pnt
1203 688 Pnt
1094 637 Pnt
1400 520 Pnt
732 850 Pnt
789 723 Pnt
1395 528 Pnt
691 947 Pnt
1446 478 Pnt
804 697 Pnt
543 1237 Pnt
574 1182 Pnt
1204 688 Pnt
1896 1284 Pnt
1781 1084 Pnt
1814 1146 Pnt
1782 1084 Pnt
807 694 Pnt
1812 1142 Pnt
1777 1075 Pnt
1495 536 Pnt
1581 661 Pnt
807 692 Pnt
689 951 Pnt
1509 555 Pnt
1195 688 Pnt
740 831 Pnt
1662 815 Pnt
1649 786 Pnt
730 855 Pnt
662 1011 Pnt
1897 1286 Pnt
1806 1131 Pnt
1601 693 Pnt
716 889 Pnt
628 1082 Pnt
1229 686 Pnt
1502 546 Pnt
1077 619 Pnt
1841 1195 Pnt
1214 688 Pnt
910 541 Pnt
991 495 Pnt
1706 921 Pnt
1890 1276 Pnt
904 550 Pnt
1690 884 Pnt
1719 950 Pnt
746 816 Pnt
1025 548 Pnt
1918 1318 Pnt
1490 529 Pnt
1322 629 Pnt
1368 571 Pnt
639 1059 Pnt
1643 773 Pnt
1452 484 Pnt
515 1281 Pnt
922 525 Pnt
1530 586 Pnt
1061 600 Pnt
1630 743 Pnt
644 1050 Pnt
548 1228 Pnt
1053 589 Pnt
1875 1251 Pnt
1817 1151 Pnt
1731 978 Pnt
1253 679 Pnt
1622 728 Pnt
594 1146 Pnt
1135 669 Pnt
985 488 Pnt
1681 863 Pnt
540 1241 Pnt
974 479 Pnt
1604 697 Pnt
579 1175 Pnt
1328 623 Pnt
1214 688 Pnt
584 1165 Pnt
752 804 Pnt
1815 1148 Pnt
1017 535 Pnt
1462 494 Pnt
1215 688 Pnt
1710 929 Pnt
1070 610 Pnt
1434 479 Pnt
1676 850 Pnt
744 823 Pnt
1246 681 Pnt
1239 683 Pnt
1720 953 Pnt
1392 533 Pnt
1870 1243 Pnt
1311 640 Pnt
1273 669 Pnt
1697 900 Pnt
1410 506 Pnt
1857 1222 Pnt
925 520 Pnt
1041 571 Pnt
1434 479 Pnt
650 1037 Pnt
1615 715 Pnt
1181 686 Pnt
1237 684 Pnt
1600 690 Pnt
1489 527 Pnt
1815 1148 Pnt
521 1272 Pnt
1557 626 Pnt
1664 822 Pnt
944 498 Pnt
1653 795 Pnt
1897 1287 Pnt
645 1048 Pnt
801 702 Pnt
1359 583 Pnt
923 523 Pnt
1813 1144 Pnt
1463 495 Pnt
1231 685 Pnt
984 487 Pnt
684 963 Pnt
1588 672 Pnt
1198 688 Pnt
1649 786 Pnt
1114 654 Pnt
1295 654 Pnt
1364 576 Pnt
1504 548 Pnt
1744 1006 Pnt
945 496 Pnt
1775 1071 Pnt
1562 633 Pnt
542 1238 Pnt
1927 1333 Pnt
590 1155 Pnt
1531 586 Pnt
1118 657 Pnt
1120 659 Pnt
614 1108 Pnt
1198 688 Pnt
782 737 Pnt
849 628 Pnt
1498 541 Pnt
862 609 Pnt
937 505 Pnt
1752 1022 Pnt
1914 1312 Pnt
1147 675 Pnt
676 981 Pnt
1569 644 Pnt
1327 624 Pnt
1515 564 Pnt
1256 677 Pnt
734 846 Pnt
1443 477 Pnt
630 1076 Pnt
1831 1177 Pnt
1423 488 Pnt
1876 1253 Pnt
943 498 Pnt
655 1027 Pnt
1783 1087 Pnt
1829 1174 Pnt
1892 1279 Pnt
1601 692 Pnt
973 478 Pnt
1034 561 Pnt
1405 513 Pnt
849 629 Pnt
1663 820 Pnt
1423 489 Pnt
1318 634 Pnt
1051 586 Pnt
1862 1231 Pnt
783 734 Pnt
1253 679 Pnt
1002 511 Pnt
1552 618 Pnt
987 490 Pnt
722 874 Pnt
977 481 Pnt
1829 1173 Pnt
1858 1223 Pnt
1224 687 Pnt
1535 592 Pnt
620 1098 Pnt
1339 610 Pnt
1342 606 Pnt
1268 672 Pnt
796 711 Pnt
871 596 Pnt
1026 548 Pnt
1518 568 Pnt
1208 688 Pnt
1633 750 Pnt
1706 920 Pnt
1154 678 Pnt
517 1277 Pnt
1765 1050 Pnt
1319 632 Pnt
1083 626 Pnt
948 492 Pnt
1230 686 Pnt
518 1277 Pnt
1191 688 Pnt
1779 1078 Pnt
1779 1079 Pnt
873 593 Pnt
640 1057 Pnt
1734 983 Pnt
1809 1137 Pnt
1164 682 Pnt
1268 671 Pnt
1456 488 Pnt
1156 679 Pnt
736 842 Pnt
716 888 Pnt
923 523 Pnt
999 507 Pnt
1134 668 Pnt
1220 687 Pnt
873 593 Pnt
1171 684 Pnt
1304 646 Pnt
1526 580 Pnt
1155 678 Pnt
1286 660 Pnt
596 1144 Pnt
636 1065 Pnt
1847 1205 Pnt
838 645 Pnt
622 1093 Pnt
1868 1240 Pnt
1647 780 Pnt
1479 515 Pnt
1225 686 Pnt
1613 711 Pnt
1285 661 Pnt
1518 568 Pnt
1436 478 Pnt
1345 602 Pnt
1203 688 Pnt
1342 607 Pnt
1021 541 Pnt
885 576 Pnt
491 1318 Pnt
810 688 Pnt
936 507 Pnt
719 882 Pnt
1343 604 Pnt
869 599 Pnt
1723 959 Pnt
1913 1311 Pnt
1550 615 Pnt
528 1260 Pnt
1361 581 Pnt
556 1214 Pnt
1129 665 Pnt
859 613 Pnt
1013 529 Pnt
1638 761 Pnt
643 1051 Pnt
1166 682 Pnt
976 480 Pnt
735 843 Pnt
1856 1220 Pnt
1828 1172 Pnt
1269 671 Pnt
932 511 Pnt
1722 958 Pnt
1194 688 Pnt
1421 491 Pnt
1536 594 Pnt
1527 581 Pnt
1425 487 Pnt
1549 613 Pnt
1682 864 Pnt
719 881 Pnt
1452 483 Pnt
567 1195 Pnt
1177 685 Pnt
693 943 Pnt
1582 663 Pnt
1257 677 Pnt
555 1216 Pnt
1501 544 Pnt
1735 986 Pnt
1619 722 Pnt
1132 667 Pnt
1221 687 Pnt
1417 496 Pnt
806 694 Pnt
574 1183 Pnt
1146 674 Pnt
1529 584 Pnt
1408 508 Pnt
499 1306 Pnt
796 710 Pnt
806 695 Pnt
519 1275 Pnt
976 480 Pnt
797 708 Pnt
1565 637 Pnt
999 506 Pnt
1917 1317 Pnt
1915 1314 Pnt
853 622 Pnt
1455 487 Pnt
921 526 Pnt
800 704 Pnt
1913 1311 Pnt
848 630 Pnt
1284 662 Pnt
603 1129 Pnt
1399 522 Pnt
820 673 Pnt
1679 858 Pnt
1606 701 Pnt
1758 1036 Pnt
1006 517 Pnt
701 924 Pnt
1721 956 Pnt
584 1164 Pnt
1243 682 Pnt
556 1214 Pnt
1469 502 Pnt
940 502 Pnt
1708 925 Pnt
1824 1163 Pnt
927 518 Pnt
982 484 Pnt
1292 656 Pnt
902 551 Pnt
521 1272 Pnt
1186 687 Pnt
1356 588 Pnt
1001 509 Pnt
826 664 Pnt
870 598 Pnt
868 601 Pnt
1237 684 Pnt
880 583 Pnt
605 1127 Pnt
628 1081 Pnt
614 1109 Pnt
1553 620 Pnt
979 482 Pnt
1047 581 Pnt
1588 672 Pnt
1578 656 Pnt
1422 490 Pnt
908 543 Pnt
571 1188 Pnt
786 728 Pnt
1606 700 Pnt
602 1131 Pnt
1245 681 Pnt
901 554 Pnt
1337 613 Pnt
734 846 Pnt
1648 784 Pnt
1192 688 Pnt
1085 627 Pnt
1725 963 Pnt
1139 671 Pnt
499 1306 Pnt
1386 543 Pnt
1096 639 Pnt
1328 623 Pnt
679 974 Pnt
1267 672 Pnt
1521 572 Pnt
1099 641 Pnt
1877 1255 Pnt
1930 1337 Pnt
1790 1100 Pnt
605 1126 Pnt
947 494 Pnt
1600 691 Pnt
930 514 Pnt
1119 658 Pnt
1079 621 Pnt
505 1297 Pnt
1308 643 Pnt
1464 496 Pnt
530 1257 Pnt
954 487 Pnt
1695 895 Pnt
1378 556 Pnt
1843 1198 Pnt
1232 685 Pnt
764 777 Pnt
753 801 Pnt
542 1237 Pnt
481 1333 Pnt
1044 576 Pnt
1122 660 Pnt
1299 650 Pnt
937 505 Pnt
1839 1191 Pnt
621 1096 Pnt
1161 680 Pnt
1216 688 Pnt
1720 952 Pnt
1858 1223 Pnt
1155 678 Pnt
479 1335 Pnt
1608 703 Pnt
899 556 Pnt
1820 1157 Pnt
1652 792 Pnt
1661 813 Pnt
1719 951 Pnt
1799 1119 Pnt
1524 576 Pnt
587 1160 Pnt
1717 946 Pnt
1663 819 Pnt
1139 671 Pnt
1400 520 Pnt
1776 1074 Pnt
1319 632 Pnt
1850 1211 Pnt
1321 630 Pnt
1204 688 Pnt
1459 491 Pnt
1573 650 Pnt
1628 739 Pnt
1711 932 Pnt
828 661 Pnt
1519 570 Pnt
1661 815 Pnt
1844 1199 Pnt
1849 1208 Pnt
1001 509 Pnt
703 920 Pnt
586 1161 Pnt
1930 1336 Pnt
890 568 Pnt
986 489 Pnt
919 528 Pnt
564 1201 Pnt
1803 1125 Pnt
1519 569 Pnt
510 1288 Pnt
896 560 Pnt
1382 549 Pnt
1785 1090 Pnt
917 532 Pnt
928 516 Pnt
574 1183 Pnt
862 610 Pnt
1436 478 Pnt
756 795 Pnt
493 1315 Pnt
1177 685 Pnt
1473 507 Pnt
1374 562 Pnt
1580 660 Pnt
1192 688 Pnt
782 737 Pnt
905 547 Pnt
712 898 Pnt
1914 1312 Pnt
1192 688 Pnt
788 724 Pnt
1256 677 Pnt
628 1082 Pnt
595 1145 Pnt
1925 1330 Pnt
1161 681 Pnt
777 746 Pnt
1687 875 Pnt
1285 661 Pnt
615 1106 Pnt
1816 1150 Pnt
1613 712 Pnt
1317 634 Pnt
1404 514 Pnt
994 499 Pnt
614 1109 Pnt
1679 856 Pnt
818 676 Pnt
1900 1291 Pnt
857 617 Pnt
564 1200 Pnt
1701 909 Pnt
1602 693 Pnt
1600 690 Pnt
1779 1078 Pnt
746 818 Pnt
1559 629 Pnt
1059 597 Pnt
823 669 Pnt
716 888 Pnt
1669 834 Pnt
1227 686 Pnt
879 585 Pnt
1170 683 Pnt
680 972 Pnt
839 644 Pnt
1882 1263 Pnt
1821 1159 Pnt
1664 822 Pnt
1889 1274 Pnt
620 1098 Pnt
1807 1133 Pnt
488 1322 Pnt
1136 669 Pnt
1496 537 Pnt
894 563 Pnt
670 993 Pnt
523 1269 Pnt
984 487 Pnt
1522 573 Pnt
666 1001 Pnt
541 1239 Pnt
1587 670 Pnt
1926 1330 Pnt
793 715 Pnt
1568 642 Pnt
1369 569 Pnt
1165 682 Pnt
1222 687 Pnt
787 726 Pnt
1589 674 Pnt
1138 670 Pnt
1885 1267 Pnt
1177 685 Pnt
1282 663 Pnt
1397 525 Pnt
1787 1094 Pnt
1306 645 Pnt
707 909 Pnt
557 1212 Pnt
1586 668 Pnt
477 1338 Pnt
1426 486 Pnt
1463 496 Pnt
1391 535 Pnt
708 908 Pnt
1223 687 Pnt
726 864 Pnt
1372 565 Pnt
536 1247 Pnt
1036 565 Pnt
743 824 Pnt
716 889 Pnt
1447 480 Pnt
1759 1038 Pnt
1152 677 Pnt
651 1035 Pnt
1571 647 Pnt
1029 553 Pnt
1703 914 Pnt
1302 648 Pnt
1748 1014 Pnt
1302 648 Pnt
1313 638 Pnt
652 1033 Pnt
497 1308 Pnt
718 884 Pnt
1527 582 Pnt
1625 733 Pnt
638 1061 Pnt
1517 566 Pnt
1460 492 Pnt
1235 684 Pnt
1599 689 Pnt
1482 519 Pnt
1347 600 Pnt
1490 529 Pnt
1450 482 Pnt
1768 1057 Pnt
1513 561 Pnt
1830 1175 Pnt
1205 688 Pnt
1259 676 Pnt
998 505 Pnt
490 1319 Pnt
1246 681 Pnt
1909 1306 Pnt
1772 1064 Pnt
1232 685 Pnt
1664 822 Pnt
622 1092 Pnt
805 697 Pnt
1430 482 Pnt
818 677 Pnt
491 1318 Pnt
662 1010 Pnt
633 1071 Pnt
620 1097 Pnt
1280 664 Pnt
654 1029 Pnt
1489 528 Pnt
778 745 Pnt
618 1101 Pnt
1111 652 Pnt
1008 520 Pnt
1041 572 Pnt
601 1133 Pnt
533 1253 Pnt
1927 1331 Pnt
638 1062 Pnt
1558 627 Pnt
1621 726 Pnt
538 1244 Pnt
995 500 Pnt
1819 1155 Pnt
962 479 Pnt
553 1219 Pnt
1809 1137 Pnt
1269 671 Pnt
1012 527 Pnt
688 955 Pnt
1858 1224 Pnt
1454 486 Pnt
1329 621 Pnt
805 695 Pnt
1783 1087 Pnt
1699 905 Pnt
486 1325 Pnt
909 542 Pnt
1380 552 Pnt
1107 648 Pnt
1171 684 Pnt
784 732 Pnt
1277 666 Pnt
1904 1298 Pnt
510 1289 Pnt
1456 488 Pnt
1497 539 Pnt
742 826 Pnt
1074 615 Pnt
1176 685 Pnt
942 499 Pnt
1073 614 Pnt
1832 1178 Pnt
1385 544 Pnt
1018 536 Pnt
1794 1108 Pnt
1697 899 Pnt
1431 481 Pnt
1326 625 Pnt
1916 1316 Pnt
1347 600 Pnt
1019 538 Pnt
1810 1138 Pnt
1420 492 Pnt
928 516 Pnt
1112 653 Pnt
523 1268 Pnt
1227 686 Pnt
1465 498 Pnt
1929 1335 Pnt
1036 564 Pnt
873 594 Pnt
1395 529 Pnt
831 655 Pnt
1214 688 Pnt
1404 514 Pnt
824 666 Pnt
597 1141 Pnt
864 606 Pnt
1310 641 Pnt
659 1017 Pnt
1182 686 Pnt
1456 487 Pnt
673 986 Pnt
815 680 Pnt
1072 613 Pnt
901 554 Pnt
819 674 Pnt
1565 637 Pnt
1775 1070 Pnt
826 664 Pnt
1662 816 Pnt
955 485 Pnt
532 1253 Pnt
766 771 Pnt
992 496 Pnt
1508 554 Pnt
1896 1286 Pnt
1398 524 Pnt
1398 524 Pnt
839 643 Pnt
584 1166 Pnt
1814 1146 Pnt
1725 964 Pnt
1158 679 Pnt
568 1193 Pnt
720 880 Pnt
1107 649 Pnt
1451 483 Pnt
1349 597 Pnt
731 854 Pnt
769 765 Pnt
1831 1177 Pnt
1916 1316 Pnt
1500 543 Pnt
1683 866 Pnt
928 517 Pnt
693 942 Pnt
807 693 Pnt
937 505 Pnt
881 582 Pnt
549 1226 Pnt
1276 667 Pnt
981 484 Pnt
1206 688 Pnt
717 886 Pnt
1044 576 Pnt
1757 1034 Pnt
1485 523 Pnt
826 664 Pnt
1248 680 Pnt
1186 687 Pnt
1673 842 Pnt
1765 1050 Pnt
1286 660 Pnt
1227 686 Pnt
1890 1275 Pnt
736 842 Pnt
1215 688 Pnt
1204 688 Pnt
1192 688 Pnt
1744 1006 Pnt
1020 539 Pnt
1059 596 Pnt
1513 560 Pnt
666 1002 Pnt
827 663 Pnt
965 478 Pnt
880 584 Pnt
1762 1044 Pnt
1079 621 Pnt
551 1222 Pnt
1191 688 Pnt
1808 1135 Pnt
1667 827 Pnt
798 707 Pnt
866 604 Pnt
497 1309 Pnt
1104 646 Pnt
625 1088 Pnt
1863 1232 Pnt
812 685 Pnt
1636 757 Pnt
1470 504 Pnt
1840 1194 Pnt
1382 550 Pnt
795 712 Pnt
1558 627 Pnt
1432 481 Pnt
902 552 Pnt
1528 582 Pnt
1509 555 Pnt
498 1308 Pnt
1908 1303 Pnt
1902 1295 Pnt
1587 671 Pnt
824 666 Pnt
659 1017 Pnt
1334 616 Pnt
1261 675 Pnt
1206 688 Pnt
1408 509 Pnt
1860 1227 Pnt
646 1046 Pnt
1752 1022 Pnt
831 656 Pnt
1475 509 Pnt
814 682 Pnt
1420 492 Pnt
686 957 Pnt
1639 763 Pnt
1061 599 Pnt
1244 682 Pnt
1121 659 Pnt
1131 666 Pnt
1017 534 Pnt
1084 627 Pnt
1667 829 Pnt
785 731 Pnt
1122 660 Pnt
602 1132 Pnt
1071 611 Pnt
1545 607 Pnt
1508 553 Pnt
1527 581 Pnt
1141 672 Pnt
1805 1128 Pnt
1295 654 Pnt
1087 630 Pnt
1204 688 Pnt
1741 1000 Pnt
1054 591 Pnt
1070 611 Pnt
771 759 Pnt
814 682 Pnt
664 1007 Pnt
729 858 Pnt
1831 1177 Pnt
1147 675 Pnt
1285 661 Pnt
702 921 Pnt
1793 1107 Pnt
1701 908 Pnt
1184 687 Pnt
1013 528 Pnt
630 1077 Pnt
1136 669 Pnt
737 839 Pnt
639 1059 Pnt
674 984 Pnt
569 1191 Pnt
1540 601 Pnt
1496 537 Pnt
1548 612 Pnt
1915 1315 Pnt
1394 531 Pnt
827 662 Pnt
947 494 Pnt
727 862 Pnt
534 1250 Pnt
1789 1099 Pnt
1628 739 Pnt
701 925 Pnt
1147 675 Pnt
1126 663 Pnt
709 906 Pnt
962 479 Pnt
511 1287 Pnt
971 478 Pnt
541 1239 Pnt
1667 829 Pnt
1899 1290 Pnt
603 1130 Pnt
1056 592 Pnt
777 747 Pnt
858 616 Pnt
1768 1056 Pnt
1595 683 Pnt
855 619 Pnt
1434 479 Pnt
862 609 Pnt
1094 637 Pnt
1073 614 Pnt
1598 687 Pnt
1829 1173 Pnt
713 897 Pnt
1446 479 Pnt
1001 510 Pnt
591 1152 Pnt
1758 1035 Pnt
1866 1237 Pnt
1245 681 Pnt
1749 1017 Pnt
1079 621 Pnt
1914 1312 Pnt
1624 732 Pnt
1005 515 Pnt
1659 809 Pnt
1706 921 Pnt
1387 542 Pnt
1042 573 Pnt
557 1212 Pnt
1619 722 Pnt
743 826 Pnt
1018 536 Pnt
1244 682 Pnt
1550 614 Pnt
1199 688 Pnt
953 487 Pnt
704 917 Pnt
782 737 Pnt
1928 1333 Pnt
1825 1166 Pnt
1827 1169 Pnt
1029 554 Pnt
777 746 Pnt
1340 609 Pnt
523 1269 Pnt
1915 1314 Pnt
687 956 Pnt
531 1256 Pnt
809 689 Pnt
1696 897 Pnt
840 642 Pnt
601 1133 Pnt
1722 958 Pnt
1747 1013 Pnt
495 1312 Pnt
544 1235 Pnt
1520 571 Pnt
1162 681 Pnt
982 485 Pnt
1347 600 Pnt
1710 929 Pnt
808 692 Pnt
1728 970 Pnt
1427 484 Pnt
686 957 Pnt
1688 878 Pnt
1166 682 Pnt
1473 508 Pnt
1050 585 Pnt
1449 480 Pnt
1175 685 Pnt
1708 924 Pnt
1769 1058 Pnt
492 1316 Pnt
1662 817 Pnt
485 1326 Pnt
1114 654 Pnt
1736 989 Pnt
1359 584 Pnt
1457 489 Pnt
705 915 Pnt
1879 1258 Pnt
1803 1126 Pnt
676 981 Pnt
581 1170 Pnt
855 620 Pnt
1016 532 Pnt
1121 660 Pnt
874 592 Pnt
1801 1121 Pnt
1628 740 Pnt
1206 688 Pnt
1318 633 Pnt
1.000 UL
LT1
400 1183 M
16 0 V
17 0 V
16 0 V
16 0 V
16 0 V
17 0 V
16 0 V
16 0 V
16 0 V
17 0 V
16 0 V
16 0 V
16 0 V
17 0 V
16 0 V
16 0 V
16 0 V
17 0 V
16 0 V
16 0 V
17 0 V
16 0 V
16 0 V
16 0 V
17 0 V
16 0 V
16 0 V
16 0 V
17 0 V
16 0 V
16 0 V
16 0 V
17 0 V
16 0 V
16 0 V
16 0 V
17 0 V
16 0 V
16 0 V
17 0 V
16 0 V
16 0 V
16 0 V
17 0 V
16 0 V
16 0 V
16 0 V
17 0 V
16 0 V
16 0 V
16 0 V
17 0 V
16 0 V
16 0 V
16 0 V
17 0 V
16 0 V
16 0 V
16 0 V
17 0 V
16 0 V
16 0 V
17 0 V
16 0 V
16 0 V
16 0 V
17 0 V
16 0 V
16 0 V
16 0 V
17 0 V
16 0 V
16 0 V
16 0 V
17 0 V
16 0 V
16 0 V
16 0 V
17 0 V
16 0 V
16 0 V
17 0 V
16 0 V
16 0 V
16 0 V
17 0 V
16 0 V
16 0 V
16 0 V
17 0 V
16 0 V
16 0 V
16 0 V
17 0 V
16 0 V
16 0 V
16 0 V
17 0 V
16 0 V
1.000 UL
LT2
400 935 M
16 0 V
17 0 V
16 0 V
16 0 V
16 0 V
17 0 V
16 0 V
16 0 V
16 0 V
17 0 V
16 0 V
16 0 V
16 0 V
17 0 V
16 0 V
16 0 V
16 0 V
17 0 V
16 0 V
16 0 V
17 0 V
16 0 V
16 0 V
16 0 V
17 0 V
16 0 V
16 0 V
16 0 V
17 0 V
16 0 V
16 0 V
16 0 V
17 0 V
16 0 V
16 0 V
16 0 V
17 0 V
16 0 V
16 0 V
17 0 V
16 0 V
16 0 V
16 0 V
17 0 V
16 0 V
16 0 V
16 0 V
17 0 V
16 0 V
16 0 V
16 0 V
17 0 V
16 0 V
16 0 V
16 0 V
17 0 V
16 0 V
16 0 V
16 0 V
17 0 V
16 0 V
16 0 V
17 0 V
16 0 V
16 0 V
16 0 V
17 0 V
16 0 V
16 0 V
16 0 V
17 0 V
16 0 V
16 0 V
16 0 V
17 0 V
16 0 V
16 0 V
16 0 V
17 0 V
16 0 V
16 0 V
17 0 V
16 0 V
16 0 V
16 0 V
17 0 V
16 0 V
16 0 V
16 0 V
17 0 V
16 0 V
16 0 V
16 0 V
17 0 V
16 0 V
16 0 V
16 0 V
17 0 V
16 0 V
0.500 UL
LTb
400 300 M
1610 0 V
0 1039 V
-1610 0 V
400 300 L
1.000 UP
stroke
grestore
end
showpage
}}%
\put(205,935){\makebox(0,0)[l]{$1/6$}}%
\put(205,1183){\makebox(0,0)[l]{$1/2$}}%
\put(1205,50){\makebox(0,0){$A\!=\!A_b\!=\!A_t$ [TeV]}}%
\put(100,819){%
\special{ps: gsave currentpoint currentpoint translate
270 rotate neg exch neg exch translate}%
\makebox(0,0)[b]{\shortstack{$|x|$}}%
\special{ps: currentpoint grestore moveto}%
}%
\put(1895,200){\makebox(0,0){ 6}}%
\put(1665,200){\makebox(0,0){ 4}}%
\put(1435,200){\makebox(0,0){ 2}}%
\put(1205,200){\makebox(0,0){ 0}}%
\put(975,200){\makebox(0,0){-2}}%
\put(745,200){\makebox(0,0){-4}}%
\put(515,200){\makebox(0,0){-6}}%
\put(350,1339){\makebox(0,0)[r]{ 1}}%
\put(350,819){\makebox(0,0)[r]{ 0.1}}%
\put(350,300){\makebox(0,0)[r]{ 0.01}}%
\end{picture}%
\endgroup
 